\documentclass[12pt]{article}
\usepackage{moreverb}

\usepackage[english]{babel}
\usepackage[latin1]{inputenc}

\usepackage{amsmath}

\usepackage{graphicx}
\usepackage{caption}
\usepackage{geometry}
\usepackage{float}
\usepackage{placeins}
\usepackage{url}
\usepackage{subfig}
\usepackage{booktabs}
\usepackage{threeparttable}
\usepackage{tikz}
\tikzstyle{decision} = [diamond, draw, fill=blue!20,
text badly centered, node distance=3cm, inner sep=0pt]
\tikzstyle{block} = [rectangle, draw, align=center, fill=blue!20,
text centered, rounded corners, minimum height=2em]
\tikzstyle{block2} = [rectangle, draw, align=center, fill=blue!20,
text centered, rounded corners, minimum height=2em, node distance=3cm]
\tikzstyle{block4} = [rectangle, draw, align=center, fill=blue!20,
text centered, rounded corners, minimum height=2em, node distance=2cm]
\tikzstyle{block3} = [align=center]
\tikzstyle{line} = [draw, -latex']
\tikzstyle{cloud} = [draw, ellipse,fill=red!20, align=center, minimum height=4em]
\tikzstyle{cloud2} = [draw, ellipse,fill=red!20, align=center, fill=red!20,
text centered, rounded corners, minimum height=3em, node distance=3cm]
\usetikzlibrary{shapes,arrows,calc, positioning}

\newcommand{\blind}{0}

\begin{document}

\def\spacingset#1{\renewcommand{\baselinestretch}%
{#1}\small\normalsize} \spacingset{1}

%%%%%%%%%%%%%%%%%%%%%%%%%%%%%%%%%%%%%%%%%%%%%%%%%%%%%%%%%%%%%%%%%%%%%%%%%%%%%%

\if0\blind
{
  \title{\bf Identifying treatment effect heterogeneity in dose-finding trials using Bayesian hierarchical models}
  \author{Marius Thomas\\
    Novartis Pharma AG, Basel, Switzerland\\
    \vspace{0.1cm}\\
    Bj\"orn Bornkamp\\ 
    Novartis Pharma AG, Basel, Switzerland\\
   \vspace{0.1cm}\\
    Katja Ickstadt\\
    TU Dortmund University, Dortmund, Germany}
  \date{}
  \maketitle
} \fi

\if1\blind
{
  \bigskip
  \bigskip
  \bigskip
  \begin{center}
    {\LARGE\bf Identifying treatment effect heterogeneity in dose-finding trials using Bayesian hierarchical models}
\end{center}
  \medskip
} \fi

\bigskip
\begin{abstract}
An important task in drug development is to identify patients, which respond
better or worse to an experimental treatment. Identifying predictive covariates, which influence the treatment effect and
can be used to define subgroups of patients, is a key aspect of this task. Analyses of treatment
effect heterogeneity are however known to be challenging, since the number of possible covariates or subgroups is
often large, while samples sizes in earlier phases of drug development are often small. In addition,
distinguishing predictive covariates from prognostic covariates, which influence the response independent
of the given treatment, can often be difficult.\\
While many approaches for these types of problems have been proposed, most of them focus
on the two-arm clinical trial setting, where patients are given either the treatment or a control.
In this paper we consider parallel groups dose-finding trials, in which patients are administered different doses
of the same treatment.\\
To investigate treatment effect heterogeneity in this setting we propose a Bayesian hierarchical dose-response model with covariate
effects on dose-response parameters. We make use of shrinkage priors
to prevent overfitting, which can easily occur, when the number
of considered covariates is large and sample sizes are small. We compare several such priors
in simulations and also investigate dependent modeling of prognostic and predictive
effects to better distinguish these two types of effects.
We illustrate the use of our proposed approach using a Phase II dose-finding trial and show how it 
can be used to identify predictive covariates and subgroups of patients with increased treatment effects.
\end{abstract}

\noindent%
\vfill

\noindent

\newpage
\spacingset{1.45} % DON'T change the spacing!

\section{Introduction}
Investigating possible heterogeneity of treatment effects is an important task in Phase II and Phase III randomized clinical trials
and can help identifying groups of patients, for which treatment effects are higher than
for the rest of the population.
By identifying such subgroups and possibly designing future clinical trials to confirm
these findings, sponsors can increase the chance of success for later trials and patients
can receive a treatment, that is particularly effective for them.
From a statistical perspective analyses of treatment effect heterogeneity with the aim
of identifying a promising subgroup are however known to be challenging. 
A large number of possible subgroups leads to
multiplicity issues, in addition treatment effect estimates in selected subgroups 
are biased (due to the selection) and sample sizes are often not adequate
for detecting treatment effect heterogeneity. A high-level overview of the
involved statistical challenges, but also opportunities is given in
\cite{rube:shen:2015}.

Commonly baseline covariates, which have been measured
before the start of the trial, are used to define subgroups. 
These covariates can be demographic, clinical or genetic and are
usually prespecified. When treatment effect
heterogeneity is investigated,
one usually aims at identifying predictive covariates, which
modify the response to the administered treatment, instead of 
just prognostic covariates which modify the response independent
of any treatment. Distinguishing prognostic and predictive effects can often
be a challenging issue \cite{sech:papa:metc:2018}. 
If covariates are identified as predictive, they can be used to define
subgroups with increased treatment effects.

A large number of methods for identifying predictive covariates and subgroups have
been proposed in recent years.
Due to their ability to handle high-order interactions and their good
interpretability, many of the proposed approaches employ tree-based
partitionings of the overall trial data \cite{su:tsai:wang:2009, fost:tayl:rube:2011, 
lipk:dmit:denn:2011, duss:vanm:2014, loh:he:man:2015, seib:zeil:hoth:2016}. 
Other statistical approaches to the problem include
Bayesian model averaging \cite{berg:wang:shen:2014} and penalized regression
\cite{imai:rat:2013, tian:ash:gent:2014}. \cite{mori:muel:2017}
recently proposed a Bayesian decision-theoretic approach. 
A recent overview paper on the topic of
subgroup identification is \cite{lipk:dmit:agos:2017}.
Most of the methods mentioned above, however, have 
been developed in the context of two-arm randomized clinical
trials that compare a new treatment against a control. In this paper we
focus on dose-finding trials, where patients are administered different
doses of the same treatment.

Dose-finding trials are an important part of drug development programs.
Trials with multiple active doses of the same treatment 
are therefore common in late-stage development, such as
Phase II dose-finding trials (see \cite{born:2017, ting:2006}
for an overview) but also Phase III trials.

\cite{thom:born:seib:2018} recently proposed a subgroup
identification approach for dose-finding trials, which makes
use of the model-based recursive partitioning algorithm \cite{zeil:hoth:horn:2008}.
This approach identifies subgroups of patients based on
parameter instabilities in $E_{max}$ models, which are commonly
used to model dose-response relationships \cite{born:2017, thom:swee:soma:2014}.
This approach shares the general advantages of recursive partitioning
methods, namely ability to deal with higher-order interactions, as well
as good interpretability. However dose-response modeling
is performed separately in each identified subgroup without
borrowing information between subgroups and thus
covariates used for splitting affect all dose-response parameters at the same time.

In this paper we propose an alternative Bayesian approach to identify 
relevant predictive covariates in dose-response trials. Our approach uses
Bayesian hierarchical non-linear dose-response models with dose-response parameters
depending on covariates. We make use of shrinkage priors, 
that support sparse solutions. These priors are used to represent the common baseline assumption, 
that many covariates are assumed to have
negligible effects on treatment effects and that strong treatment effect heterogeneity is
the exception rather than the norm. Such priors also help to
prevent overfitting in the considered setting, where commonly many covariates are 
investigated and sample sizes are quite small. 

We consider three priors: the spike-and-slab (\cite{mitc:beau:1988, geor:mccu:1993}), 
the horseshoe \cite{carv:pols:scot:2010} and the regularized
horseshoe \cite{piir:veht:2017}. The spike-and-slab is a two-component
discrete mixture prior and is often considered the gold standard for Bayesian
variable selection. The horseshoe is a continuous shrinkage prior, which
can be considered an alternative to the spike-and-slab and has shown similar or better
performance in a number of scenarios \cite{carv:pols:scot:2010, pols:scot:2011}. 
The regularized horseshoe is a recently proposed
modified version of the horseshoe, which solves MCMC convergence
issues occurring for the original horseshoe, when sampling from the posterior.

In this paper we also investigate different ways to model the relationship between prognostic and predictive effects.
While predictive
covariates are of main interest for uncovering treatment effect heterogeneity,
prognostic covariates have to be taken into consideration as well, since they can 
help explaining part of the variability in clinical response. 
Priors on prognostic and predictive effects of the same 
covariates can be modeled independently.
However there are arguments, which could be made for modeling
prognostic and predictive effects of covariates dependently.
First, from a modeling perspective, increasing the probability of prognostic
and predictive effects occuring together, helps to clearly distinguish
prognostic and predictive effects and avoids possible bias in the size
of the predictive effect. In addition, from a biological perspective, a covariate, which
has already been identified as predictive, could be considered
more likely to be prognostic as well and vice versa. 

The prognostic and predictive effects in our setting are similar to 
the concept of main effects and interactions in standard linear models.
Dependent variable selection for main effects and interactions has been
discussed in \cite{chip:1996} for discrete-mixture priors and
in \cite{grif:brow:2017} for continuous shrinkage priors. 
The above papers introduce hierarchies between main effects and interactions and reduce the chance of including
an interaction without the main effect, which reduces the prior probabilities for selecting interactions. 
Here we propose an alternative solution,
that is tailored to investigations of treatment effect heterogeneity, where the focus lies
on identifying predictive covariates, e.g. interaction. Our proposed prior distributions
increase the probability of including the prognostic effect of a covariate, when the covariate is
identified as predictive, while the marginal priors for predictive covariates are unaffected.

% Both
%of these approaches try to tackle the issue by introducing
%a hierarchy between predictors, where the amount of shrinkage
%is increasing on each level of the hierarchy and reducing the
%marginal prior probability of including effects on each level of the
%hierarchy. We show, that in our setting these approaches would drastically
%decrease the prior probability of identifying predictive effects and
%instead propose an alternative approach, which leaves
%the marginal prior distributions for the predictive effects unchanged compared
%to independent modeling.

The paper is structured as follows:
In the next section we will introduce a motivating example based on a Phase II dose-finding
study. In Section \ref{sec:methods} we will introduce our Bayesian approach, starting
with the models used and then discussing the considered shrinkage priors, as well as our proposed
solutions for modeling dependencies between prognostic and predictive effects and the identification
of subgroups. Section
\ref{sec:simstudy} contains a simulation study evaluating the
properties of the proposed method and comparing different shrinkage priors. In
Section \ref{sec:ex_analysis} the example will be revisited and
analysed to illustrate the methodology.
Conclusions and some discussion are presented in Section
\ref{sec:discussion}.

\section{A motivating example}
\label{sec:ex_intro}

In Phase II analyses are performed to identify
whether any, and if so, which baseline covariates modify the treatment
effect. The result of these analyses can be used to inform decisions 
for the design of subsequent
clinical trials.

As an example we consider data from a dose-finding trial conducted to
assess the efficacy of a new treatment for an inflammatory disease.
For reasons of confidentiality all variable names are non-descriptive
and all continuous variables have been rescaled to have mean 0 and standard
deviation 1.  We have complete
data from 270 patients, which were randomized to different dose
groups, receiving either a placebo ($n = 75$) or the new drug at dose levels
25 ($n= 54$), 50 ($n = 62$) and 100 ($n= 79$). The primary endpoint is
the change from baseline in a continuous variable. Additionally
baseline measurements of 10 covariates -- 6 of which are categorical
and 4 of which are continuous -- are available for each patient.

\begin{figure}[h!]
\centering
\includegraphics[width=0.7\textwidth]{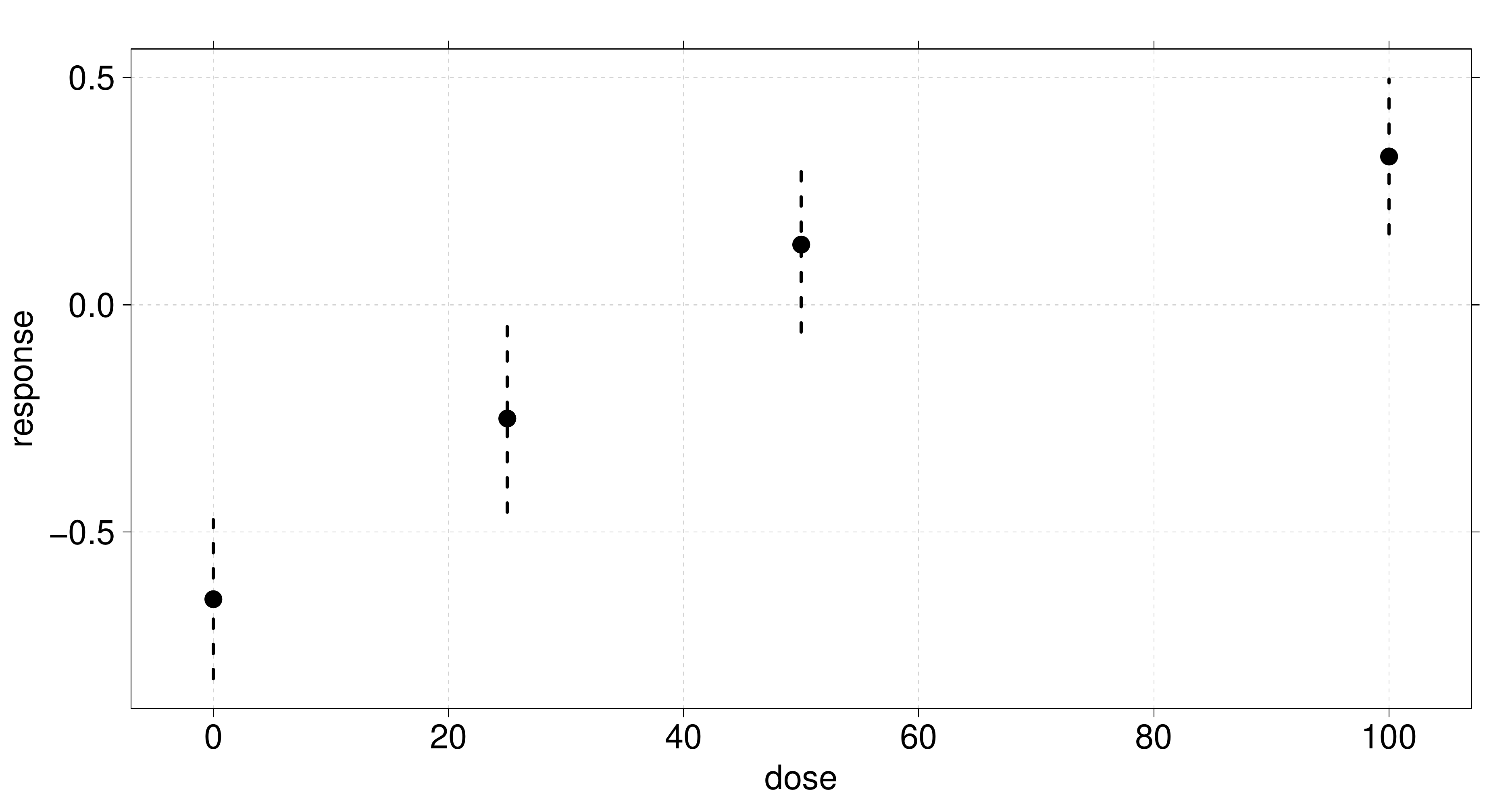}
\caption{Mean responses and $90\%$-confidence intervals for the dose-finding trial example.}
\label{fig:ex_emax}
\end{figure}

The mean responses at the dose levels in the trials along with the
confidence intervals are shown in Figure \ref{fig:ex_emax},
which suggest a clear dose-response effect. Still there is interest in
further investigation for possible predictive covariates or a subgroup with differential
treatment effect. In the next section we will develop a methodology for modelling 
the dose-response curve, but allowing for covariates to affect dose-response parameters, 
based on shrinkage priors. We will return to this example in Section \ref{sec:ex_analysis}. 

\section{Methodology}
\label{sec:methods}
We consider the situation of a clinical trial with $n$ patients, that
receive doses $d_1,...,d_n$ of a new treatment at $l$ dose levels, so
that $d_i \in \{\tilde{d_1},...,\tilde{d_l}\}$, where the lowest level
is a placebo, $\tilde{d_1} = 0$. In addition we observe outcomes
$y_1,...,y_n$.

A commonly used model for dose-response
relationships is the $E_{max}$ model, which can fit a variety
of monotonous shapes. It has been shown to be
adequate in a wide variety of real dose-response situations
\cite{thom:swee:soma:2014} and is 
non-linear in its parameters.
For the case of normally distributed
responses we can write the model as
\begin{equation}
\begin{split}
y_i \sim N(\mu_i, \sigma^2), \text{   for  } i = 1, \dots, n\\
\mu_i = E_0 + E_{max}\frac{d ^ h}{d ^ h + ED_{50} ^ h},
\end{split}
\label{eq:emaxmod}
\end{equation}
where $E_0$ can be interpreted as the mean response under placebo (d = 0),
$E_{max}$ is the maximum treatment effect as the dose goes to infinity, 
$ED_{50}$ is the dose, where a treatment effect of $0.5E_{max}$ is
reached and $h$ influences the steepness of the dose-response curve around $ED_{50}$.

For many clinical trials additional baseline covariates 
$x_i^{(1)}, \dots, x_i^{(k)}$, are measured for each patient, examples are the baseline value of
the outcome variable, clinical covariates characterizing disease severity, demographical covariates or biomarkers.
The covariates are often prespecified for possible investigation
of treatment effect heterogeneity based on clinical knowledge as well as pharmacological understanding of the drug. 
Exploratory analyses are then performed
after the trial has ended to detect possible predictive covariates, which modify the treatment effect,
and identify subgroups defined by these predictive covariates. For the dose-response situation one can include 
linear covariates on the dose-response model parameters in (\ref{eq:emaxmod}), to detect relevant prognostic and predictive covariates.

We propose to use shrinkage and variable selection priors on the effects of covariates to protect against overfitting
in these settings. As is common in the context of variable selection and penalized
regression we assume, that the covariates have been centered
and scaled to have mean 0 and standard deviation of 1. 
In the remainder of this Section 
we will generally assume, that our covariates are continuous. We discuss a possible way of dealing with categorical covariates,
when we analyze the example trial in Section \ref{sec:ex_analysis}.

\subsection{Model specification}

In the context of subgroup analyses we are interested in identifying 
covariates, which have an effect of parameters of the dose-response
model and thus influence the placebo response  $E_0$ and the treatment effect curves
of patients $E_{max}d^h/(d^h+ED50^h)$. For the dose-response parameters
we assume
\begin{equation}
\begin{split}
E_0 = \alpha_{E_0} + \beta_1 x^{(1)} + \dots + \beta_k x^{(k)}\\
E_{max} = \alpha_{E_{max}} + \gamma_1 x^{(1)} + \dots + \gamma_k x^{(k)}\\
log(ED_{50}) = \alpha_{ED_{50}} + \delta_1 x^{(1)} + \dots + \delta_k x^{(k)}.
\end{split}
\label{eq:parfunc}
\end{equation}
We assume here, that there is only limited prior knowledge about the dose-response shape
and standard deviation $\sigma$. We thus choose weakly informative, flat priors for $\alpha_{E_0}$, $\alpha_{E_{max}}$ and $\sigma$ 
and functional uniform priors for the non-linear parameters $\alpha_{ED_{50}}$ and $h$. These priors are specified
in appendix \ref{sec:app_model}.
We call the \textit{null model}, a model, where
$\boldsymbol{\beta = \gamma = \delta = 0}$. The \textit{null} model assumes the same parameters of the 
$E_{max}$ model (\ref{eq:emaxmod}) for all patients. This is similar to the model, that would be routinely
fit, when a dose-response analysis is conducted in a clinical trial, which typically however adjusts for known prognostic factors on $E_0$ as well.

\subsection{Shrinkage priors}

For the parameters $\boldsymbol{\beta, \gamma\text{  and  } \delta}$,
we could specify non-informative or weakly informative
priors, but this will lead to overfitting, since
the number of covariates can often grow quite large in the considered
settings \cite{lipk:dmit:agos:2017}. Thus it is more reasonable to choose a shrinkage prior, which
better reflects our prior beliefs of small or negligible effects for most covariates, as discussed in the Introduction.

\subsubsection{Spike-and-slab}
\label{sec:sas}
The spike-and-slab prior is often considered
to be the "gold-standard" for Bayesian variable selection \cite{mitc:beau:1988, geor:mccu:1993}. 
This prior assumes mixture distributions for the coefficients, where the first component
of the mixture is a narrow spike around zero the and the second, the "slab", is
usually a wider normal distribution. The spike-and-slab priors we use in this paper have the form
\begin{equation*}
\begin{split}
\theta \sim \lambda \cdot N(0, c^2) + (1- \lambda)\cdot \delta(0)\\
\lambda \sim Bern(p),
\end{split}
\end{equation*}
where  $\delta(0)$ is the Dirac-measure at 0, producing the spike and $c$ is the width
of the slab.
The priors we will use in this article, writing the priors in terms of a mixture over the variance, are
\begin{equation}
\begin{split}
 \beta_j \sim N(0, c_{\beta}^2{\lambda_j^{(prog)}}^2), \hspace{0.5cm} j = 1,\dots, k\\
 \gamma_j \sim N(0, c_{\gamma}^2{\lambda_j^{(pred)}}^2), \hspace{0.5cm} j = 1,\dots, k\\
 \delta_j \sim N(0, c_{\delta}^2{\lambda_j^{(pred)}}^2), \hspace{0.5cm} j = 1,\dots, k
\end{split}
\label{eq:sas}
\end{equation}
with 
\begin{equation*}
\begin{split}
\lambda_j^{(prog)} \sim Bern(\phi), \hspace{0.5cm} j = 1,\dots, k\\
\lambda_j^{(pred)} \sim Bern(\phi), \hspace{0.5cm} j = 1,\dots, k.
\label{eq:sas_lambda}
\end{split}
\end{equation*}
and priors on the slab width
\begin{equation*}
\begin{split}
c_{\beta} \sim InvGamma(0.5, 0.5)\\
c_{\gamma} \sim InvGamma(0.5, 0.5)\\
c_{\delta} \sim InvGamma(0.5, 0.5).
\end{split}
\end{equation*}

With the prior specified as above a covariate can only have a non-zero effect on both $E_{max}$ and $ED_{50}$
at the same time, which is plausible as both parameters affect the treatment effect curve.
In addition it was shown that parameters acting on $ED_{50}$ alone are hard to identify reliably [27], which is why we opt for this less complex prior.

The spike-and-slab prior prior is attractive in our setting,
since we can easily identify the important covariates based on the 
posterior distribution of the $\lambda_j$ and obtain posterior probabilities, that
the effects of certain covariates will be non-zero.
We can incorporate further prior information through the choice of the 
inclusion probabilities $\phi$, which is discussed in more detail in Section \ref{sec:hyper}. 

\subsubsection{Horseshoe}
\label{sec:hs}

%introduction, local-shrinkage prior, motivation,
% reference other horseshoe papers, 
The horseshoe prior \cite{carv:pols:scot:2010} is 
a shrinkage prior, belonging to the class of
so-called global-local shrinkage priors. For priors in these class
the variance of the coefficients is a mixture of a global and local 
component. While the global component allows shrinking spurious effects to
zero, the local component allows strong signals to persist. This makes it preferable
to other shrinkage priors, which can sometimes overshrink strong signals
\cite{carv:pols:scot:2010}.
This is an attractive property in the setting we consider, since we expect many
covariates to have no effects, however we don't want to miss strong signals
of truly predictive covariates. 

Unlike the spike-and-slab, which shrinks coefficients
either completely to zero or induces shrinkage through a normal distribution with the same variance
for all coefficients,  the horseshoe also allows for continuous shrinkage between these two extremes.
The horseshoe has shown similar or better variable selection performance than
the spike-and-slab and outperforms other continuous shrinkage priors
in many settings \cite{carv:pols:scot:2010, pols:scot:2011}, making
it a good default continuous shrinkage prior.

%The horseshoe derives its name from the prior distribution it implies
%for the so-called shrinkage factor $\kappa$, which represents the amount
%of shrinkage between 0 and 1 (where $\kappa=0$ is no shrinkage and
%$\kappa=1$ complete shrinkage). For the originally proprosed form of
%the horsehoe in \cite{carv:pols:scot:2010} $\kappa$ follows a horseshoe
%shaped $Beta(0.5, 0.5)$-distribution. 

%horseshoe priors for the coefficients from before
Using horseshoe priors for the coefficients in the dose-response scenario
results in the following prior distributions 
\begin{equation}
\begin{split}
 \beta_j \sim N(0, \tau_{\beta}^2{\lambda_j^{(prog)}}^2), \hspace{0.5cm} j = 1,\dots, k\\
 \gamma_j \sim N(0, \tau_{\gamma}^2{\lambda_j^{(pred)}}^2), \hspace{0.5cm} j = 1,\dots, k\\
 \delta_j \sim N(0, \tau_{\delta}^2{\lambda_j^{(pred)}}^2), \hspace{0.5cm} j = 1,\dots, k.
\end{split}
\label{eq:hs}
\end{equation}
As mentioned above the prior variance of the coefficients is a mixture of a local component
for that specific coefficient and a global component, which is in
our setting different between covariate effects on $E_0$, 
$ED_{50}$ and $E_{max}$. The prior distributions for
the local components are chosen as positively bounded Cauchy-
distributions,
\begin{equation*}
\begin{split}
\lambda_j^{(prog)} \sim C^{+}(0, 1),\hspace{0.5cm} j = 1,\dots, k\\
\lambda_j^{(pred)} \sim C^{+}(0, 1),\hspace{0.5cm} j = 1,\dots, k,
\label{eq:hslocal}
\end{split}
\end{equation*}
whereas the global components have distributions
\begin{equation*}
\begin{split}
\tau_{\beta} \sim C^{+}(0, \eta_{\beta} ^ 2)\\
\tau_{\gamma} \sim C^{+}(0, \eta_{\gamma} ^ 2)\\
\tau_{\delta} \sim C^{+}(0, \eta_{\delta} ^ 2) .
\end{split}
\end{equation*}
As for the spike-and-slab, local shrinkage for effects on $ED_{50}$ and $E_{max}$
is equal to reduce the number of parameters. Possible choices for the scale of the global
shrinkage parameters $\eta_{\beta}$, $\eta_{\gamma}$ and $\eta_{\delta}$
will be discussed in Section \ref{sec:hyper}.

\subsubsection{Regularized horseshoe}
\label{sec:rhs}
While the horseshoe has many attractive theoretical properties, it is
susceptible to convergence issues in practice, when using sampling methods to draw from posterior distributions 
like Stan \cite{carp:gelm:hoff:2017}. 
These issues have been discussed in \cite{piir:veht:2015} and we noticed these problems with the horseshoe as well.

To solve convergence issues, in \cite{piir:veht:2017} an extension of the horseshoe called the regularized horseshoe
was proposed. 
The regularized horseshoe prior for the coefficients in the dose-response model replaces the 
$\lambda_j^{(prog)}$ in (\ref{eq:hs}) with
\begin{equation}
\tilde{\lambda}_j^{(prog)} = \sqrt{\frac{c_{\beta} ^ 2{\lambda_j^{(prog)}} ^ 2}{c_{\beta}^2 + \tau_{\beta}^ 2{\lambda_j^{(prog)}} ^ 2}}, \hspace{0.5cm} j = 1,\dots, k,
\end{equation}
where $\lambda_j^{(prog)} \sim  C^{+}(0, 1)$ as before. Similarly the local shrinkage
components for the predictive coefficients are replaced. When there is a lot of shrinkage, 
$\tau_{\beta}^2\lambda_j^2 << c_{\beta}^2$ and $\tilde{\lambda}_j^2$ will be close to $\lambda_j ^ 2$. 
When there is only little shrinkage on the other hand the distribution of $\beta_j$ will be close to $N(0, c_{\beta}^2)$. Thus the regularized horseshoe
introduces some additional shrinkage for large effects, where the amount of shrinkage is 
determined by $c_{\beta}$ (for prognostic effects) and by $c_{\gamma}$ and $c_{\delta}$ (for predictive effects).
We choose prior distribution for $c_{\beta}$ (and same for $c_{\gamma}$ and $c_{\delta}$), 
following \cite{piir:veht:2017},
\[
c_{\beta} \sim InvGamma(2, 2).
\]

\subsection{Dependencies between prognostic and predictive effects}
% relation between prognostic and predictive effects (as title)?

The priors discussed above treat prognostic and predictive effects as independent. 
Instead we could consider priors, which introduce dependencies between
prognostic and predictive effects of covariates. In this section we propose variants of the 
spike-and-slab and the horseshoe priors, which make it unlikely to include
covariates as only predictive and instead favor covariates, to be both prognostic and predictive.
As discussed in the Introduction, there are two main arguments for using dependent priors over independent priors.
 
Firstly, by including the prognostic (main) effect, when the predictive (interaction) effect is included,
prognostic and predictive effects can be distinguished more clearly and the
of falsely identifying a prognostic effect as predictive can be reduced.

Secondly, a prior with dependencies might be more plausible from a a clinical perspective.
A covariate, that is predictive and therefore interacts with the treatment's mechanism of action
in some way, might be deemed more likely to affect the clinical response in general and therefore be prognostic
as well.  While there are certainly covariates or biomarkers, that are only predictive one might
argue that the prior probability for a covariate to be prognostic should be increased, when it is already identified
as predictive.

On the other hand, we don't want to inflate the probability to identify predictive
covariates and with it the number of false positive identifications. Thus the marginal prior probability to identify a covariate as predictive
should be same as if we were using independent priors. We can condense the above consideration into two
probability statements,
\begin{gather*}
P(x_j \text{ is prognostic}|x_j \text{ is predictive}) > P(x_j \text{ is prognostic}|x_j \text{ is not predictive})\\
P(x_j \text{ is predictive}) = p_{ind},
\end{gather*}
where $p_{ind}$ is the prior probability to include $x_j$ as predictive if we would use independent priors.
In the following we propose prior distributions for prognostic and predictive effects, which
represent these prior beliefs.

We can achieve this for the spike-and-slab by using a modified prior on the inclusion variables in (\ref{eq:sas}),
\begin{equation}
\begin{split}
\lambda_j^{(pred)} \sim Bern(\phi), \hspace{0.5cm}j = 1,\dots, k.\\
\lambda_j^{(prog)}| \lambda_j^{(pred)}  \sim Bern(\phi^{*}), \hspace{0.5cm} j = 1,\dots, k\\
\phi^{*} = 
\begin{cases}
\phi_{inc} & \text{  if  } \lambda_j^{(pred)} = 1\\
\phi & \text{  if  } \lambda_j^{(pred)} = 0,
\end{cases}
\end{split}
\end{equation}
where $\phi < \phi_{inc} \leq 1$, so the inclusion probability for prognostic effects
is increased, if the predictive effect is in the model. The choice of $\phi_{inc}$ determines
how much more likely a covariate is prognostic and predictive versus only predictive.
The resulting marginal prior distribution of $\lambda_j^{(prog)}$ is $Bern(\phi\cdot (1 - \phi + \phi_{inc}))$, while
the marginal distribution for $\lambda_j^{(pred)}$ is $Bern(\phi)$, the same as with the independent priors in Section
\ref{sec:sas}. In the remainder of this article we use $\phi_{inc} = 0.8$, so that
$P(\lambda_j^{(prog)} = 1| \lambda_j^{(pred)} = 1) = 4\cdot  P(\lambda_j^{(prog)} = 1| \lambda_j^{(pred)} = 0)$. 

Similarly, for the local shrinkage components of the horseshoe in (\ref{eq:hs}) we propose the priors
\begin{equation}
\begin{split}
\lambda_j^{(*)} \sim C^{+}(0, 1),\hspace{0.5cm} j = 1,\dots, k\\
\lambda_j^{(pred)} \sim C^{+}(0, 1),\hspace{0.5cm} j = 1,\dots, k\\
\lambda_j^{(prog)} =  max(\lambda_j^{(*)}, \lambda_j^{(pred)}),\hspace{0.5cm} j = 1,\dots, k,
\end{split}
\label{eq:hsmodlocal}
\end{equation}
which don't allow to shrink the prognostic effects more than predictive effects. For the
regularized horseshoe the same idea is used for $\tilde{\lambda}_j$.

\subsection{Practical considerations}
\subsubsection{Choice of hyperparameters}
\label{sec:hyper}
All shrinkage priors above include hyperparameters, which should be chosen in 
accordance with prior information. 
For the spike-and-slab 
prior the inclusion probabilities $\phi$ has to be specified.
To control for multiplicity in our setting the inclusion probabilities can
be lowered, when the number of considered covariates is increased, to
keep the expected number of non-zero effects constant. 
For example a choice of $\phi = 0.2$ leads to $0.2\cdot k$ covariates with non-zero effects a priori.

For the horseshoe and the regularized horseshoe the scale
of the global shrinkage parameter has to be specified. 
Originally scales of $\eta=1$ or $\eta=\sigma$ were proposed for the
global shrinkage components of the horseshoe. However \cite{piir:veht:2017} recently showed, that these
fixed choices can result in a large number of non-zero effects a priori, especially
if $k$ is large.
Instead they propose choosing the global components' scale dependent on the
problem.

In our setting we have to specify
the scales $\eta_{\beta}, \eta_{\gamma}$ and $\eta_{\delta}$.  
These global shrinkage parameters
play a similar role as the inclusion probability $\phi$ for the spike-and-slab, in that they determine the number of non-zero
coefficients expected a priori.
A possible approach to find the scales for the global shrinkage component is to use the spike-and-slab as a benchmark. By drawing samples from the prior distribution for the horseshoe with different
global scale parameters and comparing them to the spike-and-slab with the desired inclusion
probabilities we can obtain a horseshoe prior, which contains similar prior information. In principle the whole prior distribution could be compared. 
In this article we will generally try to match two prior probabilities for each set of coefficients ($\boldsymbol{
\beta, \gamma}$ and $\boldsymbol{\delta}$). The first probability represents the behavior of the prior close
to zero, while the second represents the prior behavior away from zero.  We try to minimize
\begin{equation}
g(\eta) = (P(\theta_{HS} < q_{small}|\eta) - P(\theta_{SaS, \phi} < q_{small})) ^ 2 +\\
                  (P(\theta_{HS} < q_{large}|\eta) - P(\theta_{SaS, \phi} < q_{large})) ^ 2
\label{eq:priorcrit}
\end{equation}
with regard to $\eta$, where $\theta_{HS}$ follows a horseshoe prior distribution, $\theta_{SaS, \phi}$
and has a spike-and-slab prior with inclusion probability $\phi$.
$q_{small}$ defines an interval around zero, in which any effects are considered to be negligible
for the problem at hand. On the other hand values, that are bigger than $q_{large}$ are considered to be large
effects. For example given some guesstimate for $E_{max}$, ${E_{max}}^{*}$, reasonable values could be
$q_{small} = 0.1\cdot {E_{max}}^{*}$ and $q_{large} = 1\cdot {E_{max}}^{*}$. Values for $E_0$ and $ED_{50}$
could be derived similarly.

Global scale parameters for the regularized horseshoe can be obtained with the same approach.

\subsubsection{Identifying subgroups}
\label{sec:subid}
The Bayesian modeling approach discussed above does not directly identify subgroups of patients with enhanced treatment effects.
In this Section we will outline some strategies
to identify a subgroup of patients using the output of our model. 

The Bayesian framework we are using allows us to make probability statements about individual patients' dose-response
parameters and treatment effects. We can therefore define a subgroup of patients by using the available posterior
distributions. Say for example, that we are interested in identifying a subgroup of patients with a predicted treatment effect
at a specific dose (for example the highest dose in the trial) above a certain threshold $\psi$. With our model we can estimate the treatment effects for patient
$i$ as a function of the covariates and the dose as 
\begin{equation}
\Delta_i(\boldsymbol{x_i}, d) = (\alpha_{E_{max}} + \boldsymbol{x_i^{'}\gamma})
\frac{d ^ h}{d ^ h + [exp(\alpha_{ED_{50}} + \boldsymbol{x_i^{'}\delta})]^h}.
\label{eq:trteff}
\end{equation}
We can then use the posterior distribution $P(\Delta_i(\boldsymbol{x_i}, d) | \boldsymbol{y})$ of these treatment
effect estimates to define a subgroup with an increased response as 
\begin{equation}
S = \{i \in \{1,\dots, n\}| P(\Delta_i(\boldsymbol{x_i}, d) > \psi | \boldsymbol{y}) > \omega\},
\label{eq:subid}
\end{equation}
where $\omega \in [0.5, 1]$ reflects the posterior probability for a treatment effect larger than $\psi$, that is required to place a patient in the subgroup. 
For example for $\omega = 0.5$ all patients with posterior median treatment effect above $\psi$ are placed in the subgroup. 

With the criterion above, the subgroup is chosen based on the whole covariate vector $\boldsymbol{x}$.
In many situations it may be necessary to come up with an easier subgroup description, which only depends on a small number of covariates and
is more interpretable for clinicians and patients. If only a small number of covariates
is identified as predictive, a subgroup could be defined using those covariates.
For continuous covariates a suitable cut-off has to be found.

Alternatively, a regression tree could be fit, using the posterior mean (or median) treatment effects at the highest dose or the posterior median
$E_{max}$ as the outcome and the covariates as predictors. The regression tree would then identify subgroups with different
treatment effects. A similar approach for two-arm trials is proposed in \cite{fost:tayl:rube:2011}.

\begin{figure}[b]
\centering
\resizebox{0.9\textwidth}{15cm}{%

\begin{tikzpicture}[node distance = 2cm]
% Place nodes
\node [block] (prior) {Choose prior for covariate effects};
\node [block, below left = 3.5cm and -1cm of prior] (inclusion) {Choose an inclusion probability, for example $\phi = 0.2k$};
\node [block, below right = 1.5cm and -2.5cm of prior, text width = 8cm] (guesstimate) {Choose dose-response parameter guesstimates based on clinical knowledge, previous trials, primary analysis};
\node [block, below = 0.5cm of guesstimate,  text width = 8cm] (qs) {Determine what would be large and what would be negligible effects ($q_{large}$ and $q_{small}$)};
\node [cloud, below = 1cm of qs,  text width = 5cm] (scales) { Scales for global shrinkage parameters};
\node [cloud, below left =2 cm and -2cm of inclusion,  text width = 5cm] (flat) {Uninformative priors for remaining model parameters};
\node [decision, below = 9cm of prior, text width = 3 cm] (mod) {Bayesian dose-response model};
\node [cloud, below left = 1.5cm and 3cm of mod, text width = 5cm] (trteff) {Individual treatment effect curves};
\node [cloud, below = 1.5cm of mod, text width = 5cm] (cov) {Predictive covariates};
\node [cloud, below right = 1.5cm and 3cm of mod, text width = 5 cm] (pars) {Individual dose-response parameters};
\node [decision, below = 4 cm of mod] (sub) {Subgroup identification};
\path [line] (prior) -- node[anchor = west]{horseshoe/reg. horseshoe}(guesstimate);
\path [line] (prior) -- node[anchor = east]{spike-and-slab}(inclusion);
\path [line] (qs) -- node[anchor = west]{minimize (\ref{eq:priorcrit})}(scales);
\path [line] (guesstimate) -- (qs);
\path [line] (scales) -- (mod);
\path [line] (flat) -- (mod);
\path [line] (inclusion) -- (mod);
\path [line] (mod) -- (trteff);
\path [line] (mod) -- (cov);
\path [line] (mod) -- (pars);
\path [line] (trteff) -- (sub);
\path [line] (cov) -- (sub);
\path [line] (pars) -- (sub);
\end{tikzpicture}
}
\caption{Overview over proposed procedure to identify subgroups using Bayesian hierarchical dose-response models} 
\label{flow}
\end{figure}
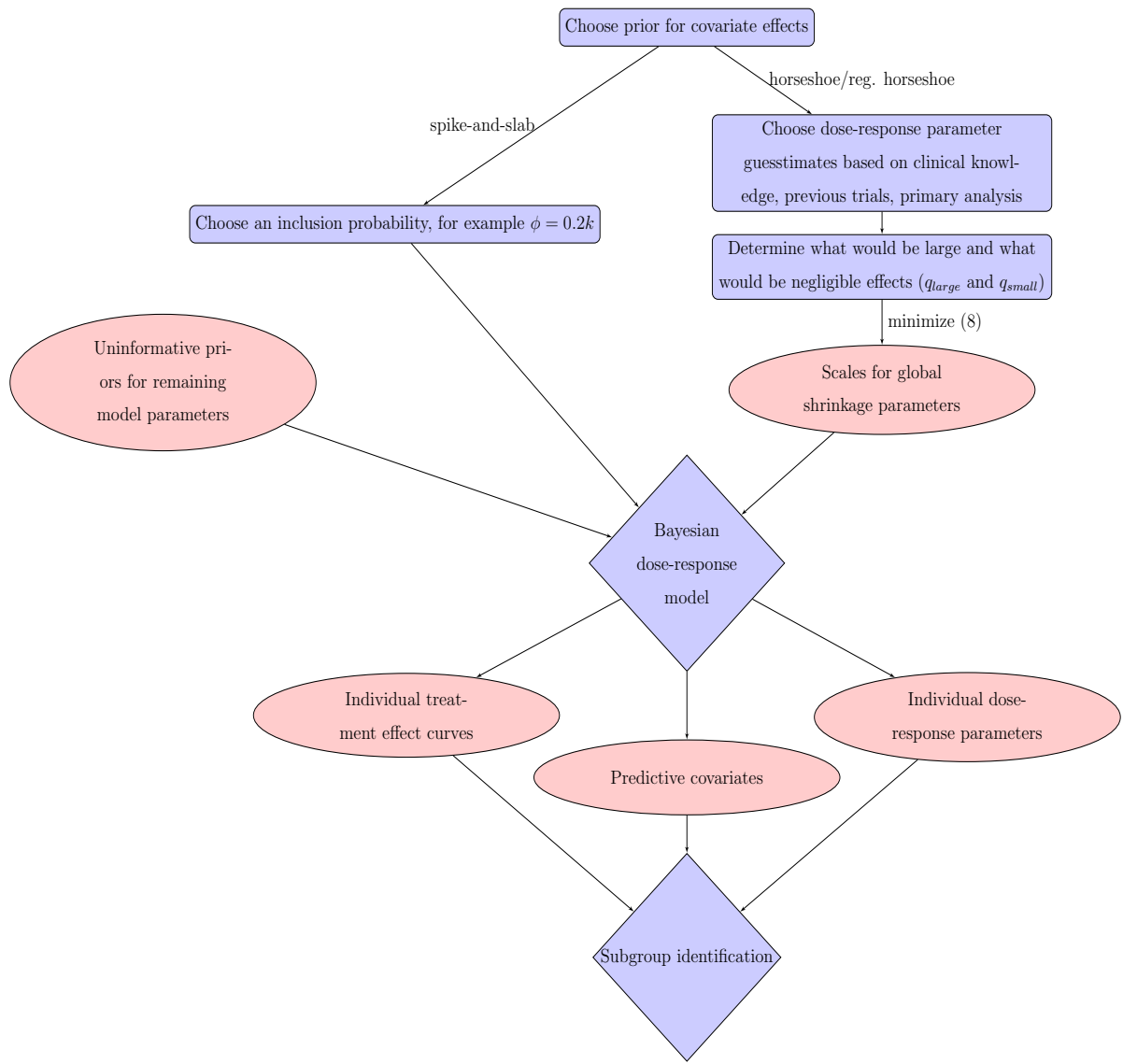

\FloatBarrier

\section{Simulation study}
\label{sec:simstudy}
In this Section we will evaluate the performance of our method using the several shrinkage priors we discussed
in Section \ref{sec:methods} using simulated dose-response trials.
The scenarios we consider are similar to the ones in \cite{thom:born:seib:2018}, but here we generally assume, that
covariate effects are linear. Some additional scenarios with non-linear functions of covariates and interactions between covariates
are discussed in Section \ref{sec:sim_nonlin}. We simulate clinical trials with $n=500$ ($n=250$) patients,
which are equally distributed across 5 dose groups with dose levels 0, 12.5, 25, 50, 100.
 We generate a vector of $k$ baseline covariates for each patient $i$
as $\boldsymbol{x_i} \sim N(\boldsymbol{0}, \boldsymbol{I_{k}})$. We consider two possible values
for the number of covariates $k$, 10 and 30.

We generate normally distributed responses from an $E_{max}$ model with $h=1$,
\begin{equation}
\begin{split}
y_i & \sim N(\mu_i, 0.25^2) \text{ i.i.d},\\
\mu_i & = E_0(\boldsymbol{x_i}) + E_{max}(\boldsymbol{x_i})\frac{d_i}{ED_{50}(\boldsymbol{x_i}) + d_i} \text{ for i=1,...n,}
\end{split}
\label{eq:datmodel}
\end{equation}
with different scenarios for covariate effects on dose-response parameters, which are summarized in Table \ref{tab:scens}.
In the first two scenarios there are no predictive covariates and all patients have the same treatment effect curve.
In the third and fourth scenario there are predictive covariates, and as a result heterogeneous treatment effect
curves for the patients. In scenario 3 and 4 $80\%$ of patients have $E_{max}$ between 0 and 0.34,
so that there are only small groups of patients with negative treatment effects or
more than double of the average treatment effect. For the $ED_{50}$ $80\%$ of patients lie between 5 and 80. 

\begin{table}[b]
\caption{Covariate effect scenarios for the simulated dose-response studies.}
\centering
\scriptsize
  \[
   \begin{array}{llll}\toprule
   \text{Scenario} & E_0(\boldsymbol{x}) & E_{max}(\boldsymbol{x}) & ED_{50}(\boldsymbol{x}) \\\midrule
    \text{1 (null)} & 1.2 & 0.17 & 20\\
    \text{2 (only prognostic)} & 1.2 + 0.1x^{(1)} + 0.1x^{(2)} + 0.05x^{(3)} & 0.17 & 20\\
    \text{3 (prognostic and predictive)} & 1.2 + 0.1x^{(1)} + 0.1x^{(2)} + 0.05x^{(3)} & 0.17 + 0.1x^{(2)} - 0.1x^{(3)} & 20 exp(-0.75x^{(2)} + 0.75x^{(3)})\\
    \text{4 (only predictive)} & 1.2 & 0.17 + 0.1x^{(2)} - 0.1x^{(3)} & 20 exp(-0.75x^{(2)} + 0.75x^{(3)})\\\bottomrule
  \end{array}
  \]
\label{tab:scens}
\end{table}
For these simulated trials we fit hierarchical dose-response models as discussed in Section \ref{sec:methods} using
different types of shrinkage priors. We also include some models without shrinkage priors for comparison.
These include a null model, which assumes constant dose-response parameters for all patients without any covariate effects,
a model with flat priors on the coefficients in (\ref{eq:parfunc}) and finally an oracle
model, which always knows the correct form of the covariate effects in Table \ref{tab:scens}. For this last model we then 
also use flat priors on all coefficients in the model. All considered methods are summarized in Table \ref{tab:competitors}.

\begin{table}[t]
\caption{Overview over different types of Bayesian dose-response models compared in the simulation study.}
\centering
  \[
   \begin{tabular}{ll}\toprule
   Abbreviation & Method \\\midrule
    null & model without covariate effects\\
    noshrink & model with flat priors\\
    oracle & model with true covariate effects\\
    hs & model with independent horseshoe shrinkage priors\\
    hs\_dep & model with dependent horseshoe shrinkage priors\\
    rhs &  model with independent reg. horseshoe shrinkage priors\\
    rhs\_dep & model with dependent reg. horseshoe shrinkage priors\\
    sas & model with independent spike-and-slab priors\\
    sas\_dep & model with dependent spike-and-slab priors\\\bottomrule
  \end{tabular}
  \]
\label{tab:competitors}
\end{table}
For the simulations we choose $\phi = \frac{2}{k}$ as the inclusion probability for the spike-and-slab prior. 
Therefore we are a priori expecting two covariates with non-zero effects, no matter how many covariates are considered in total.
For the horseshoe priors we then choose $\eta_{\beta}, \eta_{\gamma}$ and $\eta_{\delta}$, which minimize the expression (\ref{eq:priorcrit}). 
We use guesstimates ${E_0}^{*} = 1.2$, ${E_{max}}^{*} = 0.17$ and 
then follow the suggestions in section \ref{sec:hyper}, so that
$q_{small} = 0.1{E_0}^{*} = 0.12$ and $q_{large} = {E_0}^{*} = 1.2$ for $\eta_{\beta}$ and
$q_{small} = 0.1{E_{max}}^{*} = 0.017$ and $q_{large} = {E_{max}}^{*} = 0.17$ for choosing $\eta_{\gamma}$.
For $\eta_{\delta}$, we don't need any guesstimate, since coefficients for $ED_{50}$ are defined on a log-scale, and
$q_{small} = log(1.1) = 0.1$ and $q_{large} = log(2) = 0.7$ thus represent $10\%$ and
$100\%$ changes from the average $ED_{50}$.
The resulting hyperparameter values we choose for the horseshoe for different scenarios
are summarized in Table \ref{tab:simpriors}. 

\begin{table}[t]
\caption{Scales used for the global shrinkage parameter of the horseshoe and the regularized
horseshoe for  the simulation study.}
\centering
   \begin{tabular}{lrr}\toprule
    parameter & $k = 10$ & $k = 30$\\\midrule
    $\eta_{\beta}$ & 0.030 & 0.006\\
    $\eta_{\gamma}$ & 0.006 & 0.001\\
    $\eta_{\delta}$ & 0.026 & 0.005\\\bottomrule
  \end{tabular}

\label{tab:simpriors}
\end{table}
We use the following performance metrics to compare the different methods: estimation of individual patients' treatment effect curves, correct selection
of predictive covariates and identification of subgroups with increased treatment effects. In addition, in Subsection \ref{sec:sim_nonlin}
we compare the performance of the Bayesian approaches presented here to model-based recursive partitioning 
considering also non-linear functions of the covariates on the dose-response parameters. 
The results depicted in the main part of this article focus on the scenario with $n=500$ and $k=10$. Results for other considered
scenarios are similar and can be found in the Supplementary Material.

Simulations were performed using R \cite{r:2015}. For the models with Spike-and-Slab priors JAGS \cite{plum:2014} was used for sampling,
for all other models we used Stan \cite{carp:gelm:hoff:2017}.

\subsection{Estimation of individual treatment effect curves}
\label{sec:sim_trtest}
\begin{figure}[t]
\centering
\includegraphics[width=\textwidth]{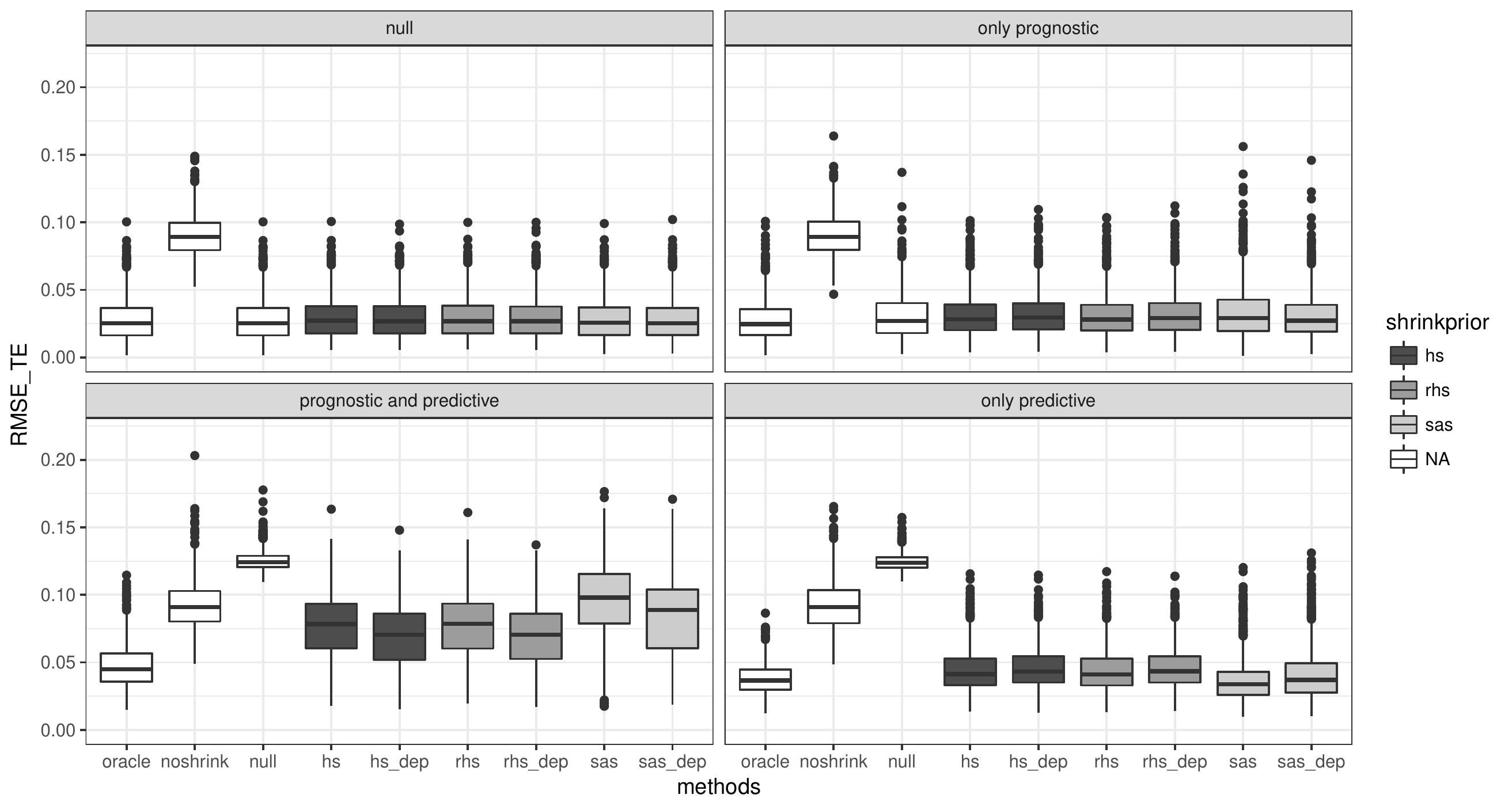}
\caption{Distribution of RMSE over 1000 simulated trials for treatment effect predictions averaged over patients and doses with
$n=500$ (100 patients per group) and $k=10$ covariates.}
\label{fig:rmse_n500_k10}
\end{figure}

With the Bayesian models we consider here we can predict an individual treatment effect curve for every patient.
In this section we compare the different models in Table \ref{tab:competitors} we consider with regard to the accuracy of their predictions.
For each patient we use the posterior mean of the treatment effect as estimated in (\ref{eq:trteff}) at the 4 active doses in the simulated
trials and then take the average of the root mean squared error (RMSE) over these doses and over
the patients. 
%
%E.g. if we denote the treatment effect for patient i
%at dose d by $\delta_i^{(d)}$ and the corresponding prediction by $\hat{\delta}_i^{(d)}$,
%we can write this metric as 
%\[
%RMSE_{\hat{\delta}} = \sqrt{\frac{1}{n}\sum\limits_{i = 1}^{n} \frac{1}{4}\sum\limits_{d\in \{12.5, 25, 50, 100\}}
%( \hat{\delta}_i^{(d)} - \delta_i^{(d)})^2}.
%\] 
The distribution of this metric over 1000 simulated clinical trials 
is visualized for the different Bayesian modeling approaches in Figure \ref{fig:rmse_n500_k10}. 

For the \textit{null}- and the \textit{only prognostic}- scenarios all patients have the same true treatment effect curves.
The shrinkage priors we consider all show the desired behaviour of shrinking all coefficients here to zero and
the resulting predictions have similar accuracy as the \textit{oracle} or the \textit{null} models, which both
assume the treatment effect is the same for all patients in these two scenarios.
The difference between the shrinkage priors with and without dependence is negligible in these two scenarios.
Only the model that assumes flat priors on the covariates performs bad here, which shows that this approach is prone to overfitting.

In the remaining scenarios with predictive effects differences between the shrinkage priors are more visible.
In scenario 3 with prognostic and predictive
effects the dependent priors (hs\_dep, rhs\_dep and sas\_dep) clearly improve the estimation of treatment effect curves compared to the independent
variants (hs, rhs, sas). Also the horseshoes priors show better performance than the spike-and-slab. In the final scenario with
only predictive effects the independent priors perform slightly better, as would be expected, since the dependent priors give this scenario
lower prior probability. Still, the difference to the dependent priors is quite small.

\subsection{Identification of predictive covariates}
\label{sec:sim_varid}
Identifying predictive covariates is a key aspect of subgroup identification. 
To determine, which covariates have effects on the dose-response parameters, we use the 50\% highest posterior density credible intervals
for the covariate effects ($\boldsymbol{\beta, \gamma}$ and $\boldsymbol{\delta}$) and assume that covariates have an effect, 
if zero is not included in the credible interval.
This criterion was suggested for Bayesian variable selection in \cite{li:lin:2010}.

Results for the variable selection performance with different shrinkage priors are summarized in Table \ref{tab:varsec_n100_k10}
for the 4 considered scenarios. In the first two scenarios all shrinkage priors correctly determine in almost all cases,
that no covariates should be selected as predictive. Covariates, that are only prognostic are slightly more often selected in 
the second scenario. The dependent priors seem to slightly reduce the selection of prognostic
covariates as predictive.

In the scenario with both prognostic and predictive effects, the dependent shrinkage priors work much better
than the independent shrinkage priors and increase the selection of all correct covariates with no visible increase in wrong
selections. In the last scenario with only predictive effects the dependent priors are sometimes slightly worse than
the independent priors, however the selection probabilities are quite similar. Interestingly the horseshoe priors seem to
be better at identifying predictive covariates on $E_{max}$, while the spike-and-slab seems better at identifying
covariates on $ED_{50}$. 
\begin{table}[ht]
\caption{Rel. frequency of covariates being selected as having effects on $E_0$, $E_{max}$ and $ED_{50}$ with $n=500$ (100 per group) and $k=10$ covariates.}
\centering
\scriptsize
\begin{tabular}{ll|rrrr|rrrr|rrrr}\toprule
& & \multicolumn{4}{c}{$E_0$} & \multicolumn{4}{c}{$ED_{50}$} & \multicolumn{4}{c}{$E_{max}$}\\
scenario & method & x1 & x2 & x3 & x4 & x1 & x2 & x3 & x4 & x1 & x2 & x3 & x4 \\ 
  \midrule
null & oracle & 0 & 0 & 0 & 0 & 0 & 0 & 0 & 0 & 0 & 0 & 0 & 0  \\ 
  & noshrink & 0.55 & 0.51 & 0.58 & 0.54 & 0.53 & 0.53 & 0.54 & 0.51 & 0.46 & 0.44 & 0.45 & 0.45 \\ 
  & hs & 0.00 & 0.00 & 0.00 & 0.00 & 0.00 & 0.00 & 0.00 & 0.00 & 0.00 & 0.00 & 0.00 & 0.00 \\ 
  & hs\_con & 0.00 & 0.00 & 0.00 & 0.00 & 0.00 & 0.00 & 0.00 & 0.00 & 0.00 & 0.00 & 0.00 & 0.00 \\ 
  & rhs & 0.00 & 0.00 & 0.00 & 0.00 & 0.00 & 0.00 & 0.00 & 0.00 & 0.00 & 0.00 & 0.00 & 0.00 \\ 
  & rhs\_dep & 0.00 & 0.00 & 0.00 & 0.00 & 0.00 & 0.00 & 0.00 & 0.00 & 0.00 & 0.00 & 0.00 & 0.00 \\ 
  & sas & 0.00 & 0.00 & 0.00 & 0.00 & 0.00 & 0.00 & 0.00 & 0.00 & 0.00 & 0.00 & 0.00 & 0.00 \\ 
  & sas\_dep & 0.00 & 0.00 & 0.00 & 0.00 & 0.00 & 0.00 & 0.00 & 0.00 & 0.00 & 0.00 & 0.00 & 0.00 \\ 
\midrule
only progn. & oracle & \it{1} & \it{1} & \it{1} & 0 & \it{0} & \it{0} & \it{0} & 0 & \it{0} & \it{0} & \it{0} & 0  \\ 
  & noshrink & \it{1.00} & \it{1.00} &  \it{0.95} & 0.56 &  \it{0.52} &  \it{0.52} &  \it{0.54} & 0.55 &  \it{0.46} &  \it{0.49} &  \it{0.46} & 0.47 \\ 
  & hs &  \it{0.99} &  \it{1.00} &  \it{0.94} & 0.02 &  \it{0.01} &  \it{0.00} &  \it{0.00} & 0.00 &  \it{0.00} &  \it{0.00} &  \it{0.01} & 0.00 \\ 
  & hs\_dep & \it{1.00} & \it{1.00} & \it{0.99} & 0.09 & \it{0.01} & \it{0.00} & \it{0.00} & 0.00 & \it{0.01} & \it{0.01} & \it{0.00} & 0.00 \\ 
  & rhs & \it{0.99} & \it{1.00} & \it{0.93} & 0.02 & \it{0.00} & \it{0.00} & \it{0.00} & 0.00 & \it{0.00} & \it{0.01} & \it{0.01} & 0.00 \\ 
  & rhs\_dep & \it{1.00} & \it{1.00} & \it{0.99} & 0.10 & \it{0.01} & \it{0.00} & \it{0.00} & 0.00 & \it{0.00} & \it{0.01} & \it{0.00} & 0.00 \\ 
  & sas & \it{0.98} & \it{0.99} & \it{0.71} & 0.00 & \it{0.01} & \it{0.01} & \it{0.03} & 0.00 & \it{0.02} & \it{0.01} & \it{0.05} & 0.00 \\ 
  & sas\_dep & \it{0.99} & \it{1.00} & \it{0.80} & 0.00 & \it{0.01} & \it{0.00} & \it{0.01} & 0.00 & \it{0.01} & \it{0.01} & \it{0.01} & 0.00\\ 
\midrule
progn. \& pred. & oracle & \it{1} & \textbf{\textit{1}} & \textbf{\textit{1}} & 0 & \it{0} & \textbf{\textit{1}} & \textbf{\textit{1}} & 0 & \it{0} & \textbf{\textit{1}} & \textbf{\textit{1}} & 0  \\ 
  & noshrink & \it{1.00} & \textbf{\textit{1.00}} & \textbf{\textit{0.86}} & 0.54 & \it{0.53} & \textbf{\textit{0.90}} & \textbf{\textit{0.89}} & 0.56 & \it{0.49} & \textbf{\textit{0.96}} & \textbf{\textit{0.96}} & 0.50 \\ 
  & hs & \it{0.99} & \textbf{\textit{0.96}} & \textbf{\textit{0.14}} & 0.02 & \it{0.00} & \textbf{\textit{0.18}} & \textbf{\textit{0.08}} & 0.00 & \it{0.01} & \textbf{\textit{0.55}} & \textbf{\textit{0.31}} & 0.00 \\ 
  & hs\_dep & \it{1.00} & \textbf{\textit{1.00}} & \textbf{\textit{0.32}} & 0.03 & \it{0.01} & \textbf{\textit{0.24}} & \textbf{\textit{0.10}} & 0.00 & \it{0.02} & \textbf{\textit{0.75}} & \textbf{\textit{0.34}} & 0.00 \\ 
  & rhs & \it{0.99} & \textbf{\textit{0.95}} & \textbf{\textit{0.14}} & 0.01 & \it{0.00} & \textbf{\textit{0.19}} & \textbf{\textit{0.09}} & 0.00 & \it{0.01} & \textbf{\textit{0.55}} & \textbf{\textit{0.29}} & 0.00 \\ 
  & rhs\_dep & \it{1.00} & \textbf{\textit{1.00}} & \textbf{\textit{0.31}} & 0.02 & \it{0.01} & \textbf{\textit{0.25}} & \textbf{\textit{0.11}} & 0.00 & \it{0.01} & \textbf{\textit{0.77}} & \textbf{\textit{0.33}} & 0.00 \\ 
  & sas & \it{0.99} & \textbf{\textit{0.84}} & \textbf{\textit{0.09}} & 0.00 & \it{0.01} & \textbf{\textit{0.29}} & \textbf{\textit{0.16}} & 0.00 & \it{0.01} & \textbf{\textit{0.45}} & \textbf{\textit{0.23}} & 0.00 \\ 
  & sas\_dep & \it{0.99} & \textbf{\textit{0.96}} & \textbf{\textit{0.18}} & 0.00 & \it{0.00} & \textbf{\textit{0.39}} & \textbf{\textit{0.19}} & 0.00 & \it{0.01} & \textbf{\textit{0.59}} & \textbf{\textit{0.28}} & 0.00 \\ 
\midrule
only pred. & oracle & 0 & \bf{0} & \bf{0} & 0 & 0 & \bf{1} & \bf{1} & 0 & 0 & \bf{1} & \bf{1} & 0  \\ 
  & noshrink & 0.52 & \bf{0.63} & \bf{0.64} & 0.56 & 0.54 & \bf{0.91} & \bf{0.89} & 0.55 & 0.46 & \bf{0.95} & \bf{0.95} & 0.47 \\ 
  & hs & 0.00 & \bf{0.03} & \bf{0.03} & 0.00 & 0.00 & \bf{0.29} & \bf{0.30} & 0.00 & 0.00 & \bf{0.91} & \bf{0.90} & 0.00 \\ 
  & hs\_dep & 0.00 & \bf{0.05} & \bf{0.05} & 0.00 & 0.00 & \bf{0.27} & \bf{0.27} & 0.00 & 0.00 & \bf{0.91} & \bf{0.90} & 0.00 \\ 
  & rhs & 0.00 & \bf{0.03} & \bf{0.03} & 0.00 & 0.00 & \bf{0.32} & \bf{0.32} & 0.00 & 0.00 & \bf{0.91} & \bf{0.91} & 0.00 \\ 
  & rhs\_dep & 0.00 & \bf{0.04} & \bf{0.04} & 0.00 & 0.00 & \bf{0.28} & \bf{0.28} & 0.00 & 0.00 & \bf{0.91} & \bf{0.91} & 0.00 \\ 
  & sas & 0.00 & \bf{0.03} & \bf{0.03} & 0.00 & 0.00 & \bf{0.76} & \bf{0.77} & 0.00 & 0.00 & \bf{0.96} & \bf{0.96} & 0.00 \\ 
  & sas\_dep & 0.00 & \bf{0.06} & \bf{0.07} & 0.00 & 0.00 & \bf{0.67} & \bf{0.69} & 0.00 & 0.00 & \bf{0.91} & \bf{0.89} & 0.00 \\ \bottomrule 
\end{tabular}
\begin{tablenotes}
\item Italic numbers indicate true prognostic covariates, bold numbers indicate true predictive covariates.
\end{tablenotes}
\label{tab:varsec_n100_k10}
\end{table}

\subsection{Subgroup identification}
\label{sec:sim_subid}

In the previous Section we have only considered the identification of single predictive covariates. However
we can also consider the identification of a subgroup with an increased treatment effect. To identify a
subgroup with an increased treatment effect we use one of the methods discussed in (\ref{sec:subid}) and estimate a subgroup
of patients with increased treatment effects as in (\ref{eq:subid}) with $\psi = 0.2$ and $\omega = 0.5$,
so that all patients with posterior median treatment effects above 0.2 at the highest dose are in the subgroup.
We can then compare this estimated subgroup $\hat{S}$ to the true subgroup of patients with such an increased effect, $S$. 
We can compare the similarity of the estimated
subgroup to the true subgroup using sensitivity ($|\hat{S}\cap S | / | S |$), specificity ($|\hat{S}^C\cap S^C | / | S ^ C |$),
positive predictive value ($|\hat{S}\cap S | / | \hat{S} |$) and negative predictive value ($|\hat{S}^C\cap S^C | / | \hat{S} ^ C |$).
Additionally we track the size of $S$ and $\hat{S}$, and if $\hat{S}$ is non-null, e.g. $\hat{S} \neq \emptyset$. The average of these metrics
over 1000 simulated clinical trials is summarized in Table \ref{tab:subid_n100_k10} in appendix \ref{sec:app_subid}.

In the first and second scenario scenario there is no subgroup with an increased treatment effect. With shrinkage
priors we still identify a subgroup $10\%$ of the time for the horseshoe and in $4\%$ for the spike-and-slab. However these
subgroups generally are small and specificity therefore is high at $97\%$. In the second scenario  the presence of prognostic covariates 
seems to lead to a subgroup being identified slightly more
often. In scenario 3 with prognostic and predictive effects sensitivity is higher for the horseshoe priors than for the spike-and-
slab, while specificity is essentially the same between methods. The dependent priors are slightly more sensitive than independent
priors. In the last scenario the independent spike-and-slab seems to perform best with regard to subgroup identification,
however all shrinkage priors show very similar performance in that scenario. 

\FloatBarrier

\subsection{Comparison to model-based recursive partitioning}
\label{sec:sim_nonlin}

Model-based recursive partitioning (\textit{mob}) has been proposed in \cite{thom:born:seib:2018} for subgroup
identification in dose-finding trials. It is of interest to investigate how the Bayesian approaches
we consider here, compare to this conceptually different approach.
Recursive partitioning methods are well known for their ability to handle interactions as well as non-linearity.
In the simulation scenarios discussed above, depicted in Table \ref{tab:scens}, we assumed, that dose-response
parameters are linear functions of the covariates. Here, to assess the robustness of the Bayesian approach described in 
this article, which always assumes linearity, in comparison to the more flexible recursive partitioning approach, we will
also consider scenarios with non-linear functions of the covariates. Apart from linear covariate effects as above, we 
additionally consider scenarios with step functions, logistic functions and interactions between covariates. We consider
these different functional forms in a scenario with prognostic and predictive effects (similar to scenario 3 in Table \ref{tab:scens}).
The different types of covariate effects considered here are summarized in Table \ref{tab:nonlin}.
\begin{table}[t]
\caption{Covariate effect scenarios for the simulated dose-response studies in Subsection \ref{sec:sim_nonlin}.}
\centering
\footnotesize
  \[
  \begin{array}{l|ll}\toprule
   \text{functional form} & E_0 & E_{max} \\\midrule
    \text{linear} & 1.2 + 0.1x^{(1)} + 0.1x^{(2)} + 0.05x^{(3)} & 0.17 + 0.1x^{(1)} - 0.1x^{(3)} \\
    \text{logistic} & 0.7 + 0.4L(x^{(1)}) + 0.4L(x^{(2)}) + 0.2L(x^{(3)}) & 0.17 + 0.34L(x^{(1)}) - 0.34L(x^{(3)}) \\
    \text{step-function} & 1.2 + 0.1I(x^{(1)}) + 0.1I(x^{(2)}) + 0.05I(x^{(3)}) & 0.17 + 0.1I(x^{(1)}) - 0.1I(x^{(3)}) \\
    \text{linear with interaction} & 1.2 + 0.1x^{(1)} + 0.1x^{(2)} + 0.05x^{(3)} + 0.2x^{(1)}x^{(2)}& 0.17 + 0.1x^{(1)} - 0.1x^{(3)} + 0.2x^{(2)}x^{(3)}\\
     & & \\\toprule
    & ED_{50} &\\\midrule
   \text{linear} & 20 \cdot exp(-0.75x^{(2)} + 0.75x^{(3)}) & \\
    \text{logistic}  & 20 \cdot exp(-2L(x^{(2)}) + 2L(x^{(3)}) & \\
     \text{step-function} & 20 \cdot exp(-0.75I(x^{(2)}) + 0.75I(x^{(3)})) & \\
     \text{linear with interaction} & 20 \cdot exp(-0.75x^{(2)} + 0.75x^{(3)} - x^{(2)}x^{(3)}) &\\\bottomrule
  \end{array}
  \]
\begin{tablenotes}
\item $L(z) := 1/(1 + exp(-2z))$ and $I(z) := I(z > 0)$.
\end{tablenotes}
\label{tab:nonlin}
\end{table}
As in the previous Sections we compare estimation of individual treatment effect curves, selection frequency of covariates as predictive 
(either on $E_{max}$ or $ED_{50}$) and subgroup identification metrics for the different Bayesian shrinkage priors and the 
\textit{mob} approach. For \textit{mob} we use the same settings as proposed
in \cite{thom:born:seib:2018}, e.g. a minimum node size of 20, significance level alpha of 0.1 and Bonferroni adjustments for multiplicity. In addition
we restrict the partitioning to covariate effects on $ED_{50}$ and $E_{max}$, since we are mostly interested in identifying predictive
covariates. 

The results for treatment effect curve estimation are depicted in Figure \ref{fig:rmse_nonlin_n500_k10} in Appendix B. As expected \textit{mob}
shows similar or better performance than the Bayesian approaches for the step-function and interaction scenario,
for which tree approaches should be particularly well suited. It is worth noting, however, that the null model
shows the best performance (apart from the oracle) in the interaction scenario. The interaction scenario
seems to be the most difficult scenario for all methods, there is also no big difference between the \textit{noshrink}
model and the models with shrinkage priors. In the other scenarios the approaches using shrinkage priors
all still work reasonably well and generally beat \textit{noshrink}- and \textit{null}-models.
\begin{table}[h!]
\caption{Rel. frequency of covariates being selected as predictive (either on $ED_{50}$ or on $E_{max}$) with $n=500$ (100 per group) and $k=10$ covariates for different
functional forms in scenarios with prognostic and predictive covariates.}
\centering
\resizebox{\textwidth}{3.5cm}{
\begin{tabular}{l|rrrr|rrrr|rrrr|rrrr}\toprule
 & \multicolumn{4}{c}{linear} & \multicolumn{4}{c}{interaction} & \multicolumn{4}{c}{logistic} & \multicolumn{4}{c}{step-function}\\
method & x1 & x2 & x3 & x4 & x1 & x2 & x3 & x4 & x1 & x2 & x3 & x4 & x1 & x2 & x3 & x4 \\
  \midrule
oracle & \it{0} & \textbf{\textit{1}} & \textbf{\textit{1}} & 0 & \it{0} & \textbf{\textit{1}} & \textbf{\textit{1}} & 0 & \it{0} & \textbf{\textit{1}} & \textbf{\textit{1}} & 0 & \it{0} & \textbf{\textit{1}} & \textbf{\textit{1}} & 0 \\ 
noshrink & \it{0.75} & \textbf{\textit{0.98}} & \textbf{\textit{0.99}} & 0.76 & \it{1.00} & \textbf{\textit{1.00}} & \textbf{\textit{1.00}} & 0.71 & \it{0.76} & \textbf{\textit{0.90}} & \textbf{\textit{0.89}} & 0.71 & \it{0.71} & \textbf{\textit{0.84}} & \textbf{\textit{0.83}} & 0.72 \\
hs & \it{0.01} & \textbf{\textit{0.56}} & \textbf{\textit{0.32}} & 0.01 & \it{1.00} & \textbf{\textit{1.00}} & \textbf{\textit{0.98}} & 0.01 & \it{0.03} & \textbf{\textit{0.33}} & \textbf{\textit{0.03}} & 0.00 & \it{0.02} & \textbf{\textit{0.52}} & \textbf{\textit{0.01}} & 0.01 \\ 
hs\_dep & \it{0.03} & \textbf{\textit{0.77}} & \textbf{\textit{0.35}} & 0.00 & \it{1.00} & \textbf{\textit{1.00}} & \textbf{\textit{0.98}} & 0.01 & \it{0.03} & \textbf{\textit{0.36}} & \textbf{\textit{0.03}} & 0.00 & \it{0.03} & \textbf{\textit{0.42}} & \textbf{\textit{0.01}} & 0.00 \\ 
rhs & \it{0.01} & \textbf{\textit{0.56}} & \textbf{\textit{0.30}} & 0.00 & \it{1.00} & \textbf{\textit{1.00}} & \textbf{\textit{0.99}} & 0.01 & \it{0.03} & \textbf{\textit{0.32}} & \textbf{\textit{0.04}} & 0.00 & \it{0.02} & \textbf{\textit{0.41}} & \textbf{\textit{0.01}} & 0.01 \\ 
rhs\_dep & \it{0.02} & \textbf{\textit{0.79}} & \textbf{\textit{0.35}} & 0.00 & \it{1.00} & \textbf{\textit{1.00}} & \textbf{\textit{0.99}} & 0.01 & \it{0.03} & \textbf{\textit{0.35}} & \textbf{\textit{0.04}} & 0.00 & \it{0.03} & \textbf{\textit{0.33}} & \textbf{\textit{0.01}} & 0.00 \\ 
sas & \it{0.01} & \textbf{\textit{0.46}} & \textbf{\textit{0.24}} & 0.00 & \it{0.98} & \textbf{\textit{0.98}} & \textbf{\textit{0.95}} & 0.00 & \it{0.08} & \textbf{\textit{0.41}} & \textbf{\textit{0.04}} & 0.00 & \it{0.06} & \textbf{\textit{0.56}} & \textbf{\textit{0.01}} & 0.00 \\ 
sas\_dep & \it{0.01} & \textbf{\textit{0.60}} & \textbf{\textit{0.28}} & 0.00 & \it{1.00} & \textbf{\textit{1.00}} & \textbf{\textit{0.96}} & 0.00 & \it{0.03} & \textbf{\textit{0.32}} & \textbf{\textit{0.02}} & 0.00 & \it{0.02} & \textbf{\textit{0.41}} & \textbf{\textit{0.00}} & 0.00 \\  
mob & \it{0.23} & \textbf{\textit{0.96}} & \textbf{\textit{0.30}} & 0.10 & \it{0.32} & \textbf{\textit{0.77}} & \textbf{\textit{0.54}} & 0.11 & \it{0.14} & \textbf{\textit{0.87}} & \textbf{\textit{0.15}} & 0.08 & \it{0.08} & \textbf{\textit{0.65}} & \textbf{\textit{0.10}} & 0.05 \\\bottomrule 
\end{tabular}}
\begin{tablenotes}
\item \small Italic numbers indicate true prognostic covariates, bold numbers indicate true predictive covariates.
\end{tablenotes}
\label{tab:varsec_nonlin}
\end{table}
Simulation results for selection frequency of covariates as predictive are summarized in Table \ref{tab:varsec_nonlin}.
When using \textit{mob} we consider a covariate to be selected as predictive, if it is at any point used for a split in the tree.
\textit{Mob} generally seems to select more covariates (right or wrong ones) than the Bayesian approaches with shrinkage priors.
The only exception seems to be the scenario with interactions, where the Bayesian approaches essentially always select the
correct predictive covariates, but also the first covariate, which is only prognostic. In the interaction scenario \textit{mob} also
identifies this covariate more often, however the rate of identification is much lower than for the Bayesian approaches.  
Overall it seems that \textit{mob} is more robust with regard to possible functional forms as the results seem quite similar for all considered scenarios.
The Bayesian approaches work better than \textit{mob} in the linear scenario, however they miss more predictive covariates in
the logistic and step-function scenarios and seem to be unable to distinguish prognostic and predictive effects in the interaction
scenario.

The subgroup identification performance is summarized in Table \ref{tab:subid_nonlin_n100_k10} in appendix \ref{sec:app_nonlin}. \textit{Mob} generally identifies larger subgroups and has 
similar or better sensitivity than the Bayesian approaches. On the other hand the specificity of \textit{mob} is generally lower.  

\section{Analysis of the example trial}
\label{sec:ex_analysis}
We now come back to our example trial and use the shrinkage approach discussed above on this dataset.
As discussed in Section \ref{sec:ex_intro} we have 10 covariates in total, which we consider here as possible predictive
covariates. 6 of the covariates in this dataset are categorical. Dealing with categorical covariates introduces an additional challenge in the context of variable
selection. Dummy-coding is commonly used for categorical covariates, therefore for a covariate
with $Z$ categories we introduce $Z-1$ dummy variables. In our model these dummy variables are then considered to be separate binary covariates,
however it does not make sense to shrink them completely independently, since we want to shrink the effects of the underlying covariates, not
of the dummy variables used for coding. We therefore use the same local shrinkage components for all dummy variables
belonging to the same covariate.

To get an overall impression of the data without covariate effects, we fit a non-Bayesian sigmoid $E_{max}$ model to the data
assuming no covariate effects. The results confirm the relatively clear dose-response trend from the visual
inspection of the data. We obtain ML-estimates $E_0 = -0.65$, $E_{max} = 1.04$, $ED_{50} = 30.90$ and $h = 2.27$.

For the exploratory analysis for possible predictive covariate effects and subgroups we only consider the dependent regularized horseshoe, 
since this prior seemed to show good performance in the simulation study in the previous Section.
As discussed in Section \ref{sec:hs}, we have to specify the scale for the global shrinkage component. 
We can use the ML-estimates above and as in Section \ref{sec:simstudy} consider $10\%$ changes and $100\%$ 
changes in the dose-response parameters as small and large effects.
Minimizing criterion (\ref{eq:priorcrit}), we obtain global scale parameters $\tau_{\beta} = 0.020$, 
$\tau_{\gamma} = 0.030$ and $\tau_{\delta} = 0.039$. 
We then fit a a Bayesian sigmoid $E_{max}$-model as in (\ref{eq:emaxmod}) with covariate effects as in (\ref{eq:parfunc}) using
the dependent regularized horseshoe priors. 

\begin{figure}[t!]
\centering
\includegraphics[width=0.8\textwidth]{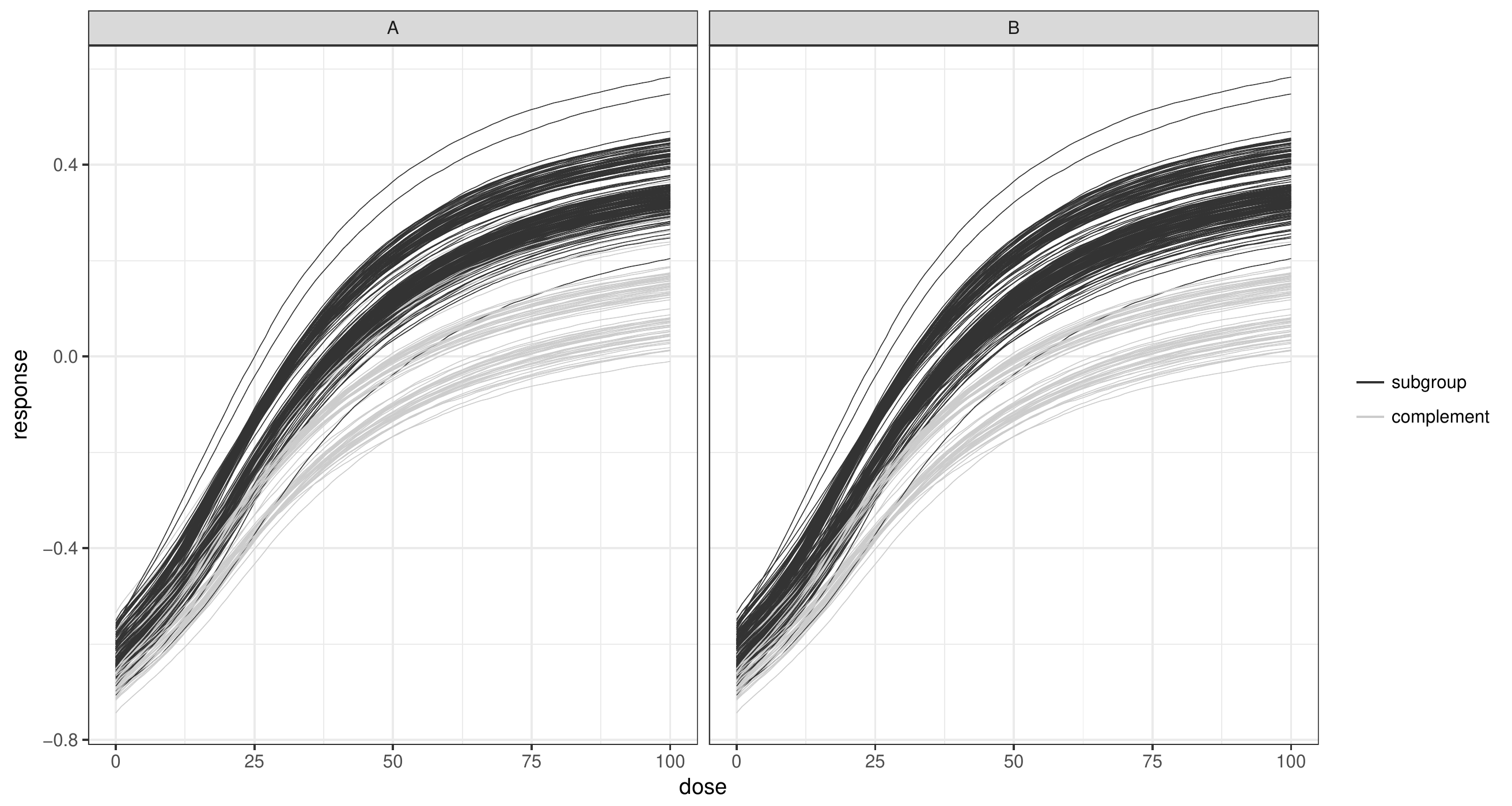}
\caption{Individual predicted dose-response curves for patients in subgroup and complement, which is identified using either
the individual $E_{max}$ (A) or the predictive covariate $x_7$ (B).}
\label{fig:exsub}
\end{figure}

We can get an idea of which covariates are most likely to be predictive by considering the size of the local shrinkage coefficients
$\boldsymbol{\lambda^{(pred)}}$. Going by the median local shrinkage components the most important covariates are $x^{(7)}$, $x^{(9)}$
and $x^{(6)}$, the first two of which are binary, while the last is continuous. 
The summary statistics for the coefficients for these covariates as well as for other important parameters of the model are given in Table \ref{tab:ex_summ}.
Looking at the coefficients for the three important covariates, almost all of the posterior medians are close to zero. The only exception is $\gamma_7$, which
describes the effect of the binary covariate $x_7$ on $E_{max}$. The $50\%$ credible interval for this coefficient however still includes zero.
The conclusion here seems to be, that $x_7$ might be predictive on $E_{max}$, however the trial at hand is likely to small to make definite
conclusions. This finding could be further investigated in another, larger trial.
\begin{table}[h]
\caption{Posterior summaries for selected parameters of the fitted dose-response model}
\centering
\footnotesize
$\begin{array}{lrrrrrrrr}\toprule
\text{parameter} & \text{mean} & \text{sd} & 2.5\% & 25\% & 50\% & 75\% & 97.5\% \\ 
  \midrule
  \alpha_{E0} & -0.63 & 0.11 & -0.84 & -0.70 & -0.63 & -0.56 & -0.42 \\ 
  \alpha_{Emax} & 1.11 & 0.29 & 0.67 & 0.90 & 1.06 & 1.27 & 1.78 \\ 
  exp(\alpha_{ED50}) & 34.58 & 14.00 & 11.68 & 25.12 & 31.88 & 41.90 & 68.47 \\ 
  h & 2.68 & 1.96 & 0.65 & 1.28 & 2.01 & 3.45 & 8.17 \\ 
  \sigma & 0.91 & 0.04 & 0.83 & 0.88 & 0.91 & 0.93 & 0.99 \\ 
  \lambda_1^{(pred)} & 1.68 & 6.88 & 0.03 & 0.34 & 0.78 & 1.65 & 8.05 \\ 
 \lambda_2^{(pred)} & 3.98 & 77.21 & 0.03 & 0.35 & 0.84 & 1.79 & 9.79 \\ 
  \lambda_3^{(pred)} & 0.91 & 1.13 & 0.03 & 0.27 & 0.59 & 1.14 & 3.83 \\ 
 \lambda_4^{(pred)} & 1.58 & 2.80 & 0.03 & 0.35 & 0.82 & 1.73 & 7.75 \\ 
  \lambda_5^{(pred)} & 1.90 & 6.08 & 0.04 & 0.36 & 0.83 & 1.81 & 10.19 \\ 
  \lambda_6^{(pred)} & 3.35 & 38.47 & 0.04 & 0.41 & 0.99 & 2.24 & 15.47 \\ 
  \lambda_7^{(pred)} & 8.32 & 64.87 & 0.07 & 0.91 & 2.54 & 6.02 & 39.61 \\ 
  \lambda_8^{(pred)} & 1.56 & 3.25 & 0.03 & 0.35 & 0.78 & 1.65 & 7.49 \\ 
 \lambda_9^{(pred)}& 3.06 & 15.20 & 0.05 & 0.49 & 1.17 & 2.74 & 14.16 \\ 
  \lambda_{10}^{(pred)} & 0.89 & 1.10 & 0.03 & 0.26 & 0.58 & 1.13 & 3.63 \\ 
  \beta_7 & 0.03 & 0.06 & -0.04 & -0.00 & 0.01 & 0.06 & 0.18 \\ 
  \beta_9 & 0.02 & 0.04 & -0.03 & -0.00 & 0.00 & 0.03 & 0.14 \\ 
  \beta_{6} & 0.01 & 0.04 & -0.04 & -0.00 & 0.00 & 0.01 & 0.12 \\ 
 \gamma_7 & 0.10 & 0.13 & -0.03 & 0.00 & 0.04 & 0.17 & 0.43 \\ 
 \gamma_9 & 0.02 & 0.07 & -0.06 & -0.00 & 0.00 & 0.04 & 0.20 \\ 
  \gamma_{6} & -0.03 & 0.08 & -0.29 & -0.03 & -0.00 & 0.00 & 0.05 \\ 
 \delta_7 & 0.02 & 0.36 & -0.40 & -0.03 & -0.00 & 0.02 & 0.53 \\ 
  \delta_9 & -0.05 & 0.14 & -0.44 & -0.04 & -0.00 & 0.00 & 0.08 \\ 
 \delta_6 & 0.02 & 0.11 & -0.10 & -0.01 & 0.00 & 0.02 & 0.28 \\\bottomrule 
\end{array}$
\label{tab:ex_summ}
\end{table}
 
Based on the above finding we would likely conclude in a real-life setting, that there is not enough evidence for a possible subgroup.
However we will continue with a possible subgroup identification procedure based on the above models for the purposes of this example. 
As discussed in Section \ref{sec:subid} there are several options, that can be considered to define a subgroup of patients
with a high treatment response.
For example we could consider all patients with posterior median $E_{max}$ above 1.04 (the ML-estimate for the primary analysis)
as a possible subgroup. Alternatively we could consider, that $x_7$ is a binary covariate, which has been identified as possibly predictive and could
be used to define a subgroup. Figure \ref{fig:exsub} shows the individual predicted dose-response curves for patients in and out of
the subgroups identified in these ways.  The resulting subgroups are almost identical, which further confirms, that $x_7$ is the only
important predictive covariate.

\FloatBarrier
\section{Conclusions and discussion}
\label{sec:discussion}

In this paper we proposed a Bayesian approach for investigating treatment effect heterogeneity
in dose-response trials. Our approach uses shrinkage priors to take into account, that commonly 
little systematic heterogeneity of treatment effects is expected. We proposed ways to incorporate prior beliefs and control multiplicity in a Bayesian way through the hyperparameters
of the prior distributions on prognostic and predictive effect and presented ways to model dependencies between prognostic and predictive effects
of the same covariate. Our methods can deal with continuous, as well as categorical covariates as showed in Section \ref{sec:ex_analysis}.
While we focused on normally distributed outcomes, the presented Bayesian model should be extendable to other types
of outcomes as well.

Our simulation studies in Section \ref{sec:simstudy} show, that the considered shrinkage priors all have the desired behaviour of
reducing the number of false positive identifications of covariates as predictive. This can be seen directly in Section \ref{sec:sim_varid}, where
covariates, that are not predictive are almost never selected as such.
Based on the results in Section \ref{sec:sim_trtest} this also results in better estimation of individual treatment effect curves.
In scenarios with predictive effects some differences were visible between the priors with regard to the identification performance.
Our results suggest, that the horseshoe priors perform better than the spike-and-slab priors in the scenario with prognostic and predictive
effects, both with regard to variable selection and treatment effect estimation. These properties of the priors are also reflected in the 
performance with regard to subgroup identification in Section \ref{sec:sim_subid}. Differences in performance between horseshoe and
regularized horseshoe are negligible, however we noticed the improved sampling properties for the regularized horseshoe, which shows a 
much smaller number of divergences, when using Stan. 

As expected priors, which model prognostic and predictive effects dependently, gave good performance improvements in the scenarios with prognostic and
predictive effects. Additionally the performance in the other scenarios was similar to the independent priors.
The dependent priors  therefore seemed to show the desired behaviour of improving the ability of the method to detect predictive covariates, that are also prognostic, while only minimally
increasing the rate of false identifications for covariates, which are only prognostic. Based on our results 
we would therefore recommend dependent priors over independent priors, when the main interest of the analysis is
the identification of predictive covariates.

We compared the performance of the Bayesian approaches discussed in this
article to model-based recursive partitioning \cite{thom:born:seib:2018}, 
a recently proposed method for identifying subgroups with heterogeneous treatment effects in dose-findings trials.
in Section \ref{sec:sim_nonlin}. \textit{mob} showed higher rates of both true and false positive selections in the scenarios
we considered. In addition \textit{mob}'s performance was quite robust to the different functional forms we considered, whereas the
performance of the approaches presented in this article was more dependent on the functional form of the covariates, 
due to assuming linearity of the covariates. However, the Bayesian models generally better estimate individual treatment
effect curves (see Figure \ref{fig:rmse_nonlin_n500_k10}
in appendix 
\ref{sec:app_nonlin}), since they borough information across subgroups and prefer sparse solutions.
Based on our results it is hard to generally recommend one method over the other, since the two approaches are conceptually
different and both have their advantages and disadvantages. In addition to the differences in operating characteristics, 
\textit{mob} also requires less input from the user and the tree output is
easily interpretable, while the Bayesian models will generally require more input from the user, but also
provide a richer output with proper statements of uncertainty for all parameters and quantities of interest.

This leads us to the limitations of the discussed approaches. Firstly we only consider linear functions of the covariates in our models,
which also do not include any covariate-covariate interactions. The results of this can be seen in Section \ref{sec:sim_nonlin} and
especially in Table \ref{tab:varsec_nonlin}: When
the covariates do not affect the dose-response parameters linearly the identification performance for our methods worsens.
In the scenario with interaction terms our method seems to have trouble distinguishing prognostic and predictive covariates. However
this would likely be a challenging scenario for any method, since the prognostic covariate $x_1$ has a (prognostic) interaction with the predictive covariate $x_2$.
\textit{mob} also shows increased identifications of $x_1$ as predictive in this scenario.
Allowing for more complex functional forms, for example using basis function expansions in the Bayesian framework with shrinkage priors,
could be a possibility for further research.

\newpage

\section*{Acknowledgements}
\small
 This work was supported by funding from the European Union's Horizon 2020 research and innovation programme under the Marie Sklodowska-Curie grant agreement No 633567 and by the Swiss State Secretariat for Education, Research and Innovation (SERI) under contract number 999754557 . The opinions expressed and arguments employed herein do not necessarily reflect the official views of the Swiss Government.\\

\begin{minipage}{.5\textwidth}
\includegraphics[scale = 0.03]{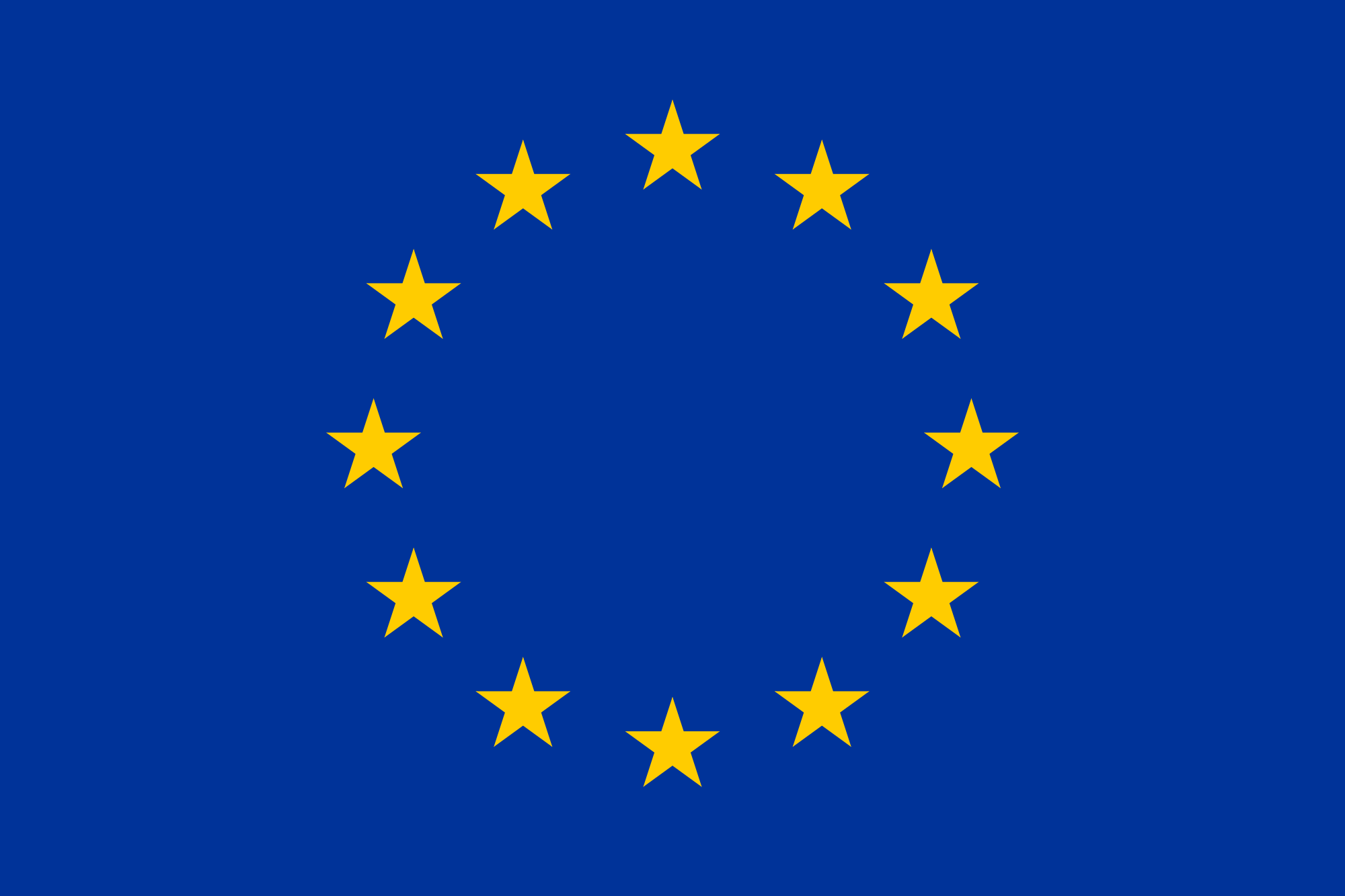}
\end{minipage}%
\begin{minipage}{.5\textwidth}
\includegraphics[scale = 1]{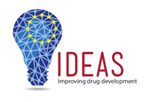}
\end{minipage}

\section*{Appendix}
\appendix

\section{Full model specification}
\label{sec:app_model}

To complete the priors for model (\ref{eq:parfunc}) we choose
\begin{gather*}
 \alpha_{E_0} \sim N(0, \sigma_{E_0}^2)\\
 \alpha_{E_{max}} \sim N(0, \sigma_{E_{max}}^2)\\
 \alpha_{ED_{50}} = log(\nu \cdot d_{max})\\
 h = 0.5 + 9.5 \cdot \xi\\
 \sigma \sim InvGamma(0.01, 0.01).
\end{gather*}
Here $\nu$ and $\xi$ are hyperparameters, while $\sigma_{E_0}$
and  $\sigma_{E_{max}}$ are fixed. Since we assume limited prior knowledge we would
choose flat priors for $E_0$ and $E_{max}$ and choose large values for
$\sigma_{E_0}$ and $\sigma_{E_{max}}$.
For the hyperparameters we choose prior distributions
\begin{gather*}
 \nu\sim Beta(0.82, 3.5)\\
 \xi \sim Beta(0.93, 1.4).
\end{gather*}
These beta distributions lead to functional uniform priors on the non-linear parameters of the 
model, $ED_{50}$ and $h$, which result in uniform
distributions over the possible functional shapes of the dose-response curve  (see \cite{born:2014} for details).

\section{Simulation results for subgroup identification}
\label{sec:app_subid}
\begin{table}[h]
\caption{Subgroup identification metrics for scenarios with $n=500$ (100 per group) and $k=10$ covariates.}
\tiny
\centering
\begin{tabular}{llrrrrrrr}\toprule
scenario & method & $|S|$ & $|\hat{S}|$ & non.null & sens & spec & ppv & npv \\ 
  \midrule
null & oracle & 0.00 & 14.00 & 0.03 &  & 0.97 & 0.00 & 1.00 \\ 
  & noshrink & 0.00 & 41.90 & 0.96 &  & 0.92 & 0.00 & 1.00 \\ 
  & hs & 0.00 & 14.85 & 0.10 &  & 0.97 & 0.00 & 1.00 \\ 
  & hs\_dep & 0.00 & 15.16 & 0.10 &  & 0.97 & 0.00 & 1.00 \\ 
  & rhs & 0.00 & 14.73 & 0.10 &  & 0.97 & 0.00 & 1.00 \\ 
  & rhs\_dep & 0.00 & 14.91 & 0.10 &  & 0.97 & 0.00 & 1.00 \\ 
  & sas & 0.00 & 14.01 & 0.04 &  & 0.97 & 0.00 & 1.00 \\ 
  & sas\_dep & 0.00 & 14.75 & 0.04 &  & 0.97 & 0.00 & 1.00 \\ 
\midrule 
 only progn. & oracle & 0.00 & 11.00 & 0.02 &  & 0.98 & 0.00 & 1.00 \\ 
  & noshrink & 0.00 & 43.73 & 0.97 &  & 0.91 & 0.00 & 1.00 \\ 
  & hs & 0.00 & 16.51 & 0.21 &  & 0.97 & 0.00 & 1.00 \\ 
  & hs\_dep & 0.00 & 18.08 & 0.26 &  & 0.96 & 0.00 & 1.00 \\ 
  & rhs & 0.00 & 16.53 & 0.22 &  & 0.97 & 0.00 & 1.00 \\ 
  & rhs\_dep & 0.00 & 18.17 & 0.26 &  & 0.96 & 0.00 & 1.00 \\ 
  & sas & 0.00 & 22.12 & 0.21 &  & 0.96 & 0.00 & 1.00 \\ 
  & sas\_dep & 0.00 & 16.78 & 0.13 &  & 0.97 & 0.00 & 1.00 \\
\midrule
progn. \& pred. & oracle & 175.29 & 150.00 & 1.00 & 0.78 & 0.96 & 0.93 & 0.90 \\ 
  & noshrink & 175.29 & 107.94 & 1.00 & 0.51 & 0.94 & 0.84 & 0.79 \\ 
  & hs & 175.29 & 138.05 & 0.93 & 0.62 & 0.91 & 0.84 & 0.84 \\ 
  & hs\_dep & 175.29 & 149.93 & 0.98 & 0.68 & 0.91 & 0.84 & 0.86 \\ 
  & rhs & 175.29 & 139.04 & 0.93 & 0.62 & 0.91 & 0.84 & 0.84 \\ 
  & rhs\_dep & 175.29 & 149.99 & 0.98 & 0.68 & 0.91 & 0.84 & 0.86 \\ 
  & sas & 175.29 & 118.19 & 0.80 & 0.50 & 0.91 & 0.80 & 0.79 \\ 
  & sas\_dep & 175.29 & 132.30 & 0.86 & 0.58 & 0.91 & 0.81 & 0.82 \\ 
\midrule
only pred. & oracle & 175.60 & 154.13 & 1.00 & 0.82 & 0.97 & 0.95 & 0.92 \\ 
  & noshrink & 175.60 & 107.77 & 1.00 & 0.51 & 0.95 & 0.85 & 0.78 \\ 
  & hs & 175.60 & 167.12 & 1.00 & 0.84 & 0.94 & 0.90 & 0.92 \\ 
  & hs\_dep & 175.60 & 167.23 & 1.00 & 0.83 & 0.94 & 0.90 & 0.92 \\ 
  & rhs & 175.60 & 167.91 & 1.00 & 0.84 & 0.94 & 0.90 & 0.92 \\ 
  & rhs\_dep & 175.60 & 167.64 & 1.00 & 0.83 & 0.93 & 0.90 & 0.92 \\ 
  & sas & 175.60 & 168.73 & 1.00 & 0.86 & 0.95 & 0.92 & 0.93 \\ 
  & sas\_dep & 175.60 & 165.19 & 1.00 & 0.84 & 0.94 & 0.91 & 0.92 \\\bottomrule 
\end{tabular}
\label{tab:subid_n100_k10}
\end{table}

\FloatBarrier

\newpage
\section{Comparison to model-based recursive partitioning: further simulation results}
\label{sec:app_nonlin}
\begin{figure}[h!]
\centering
\scriptsize
\includegraphics[width=\textwidth]{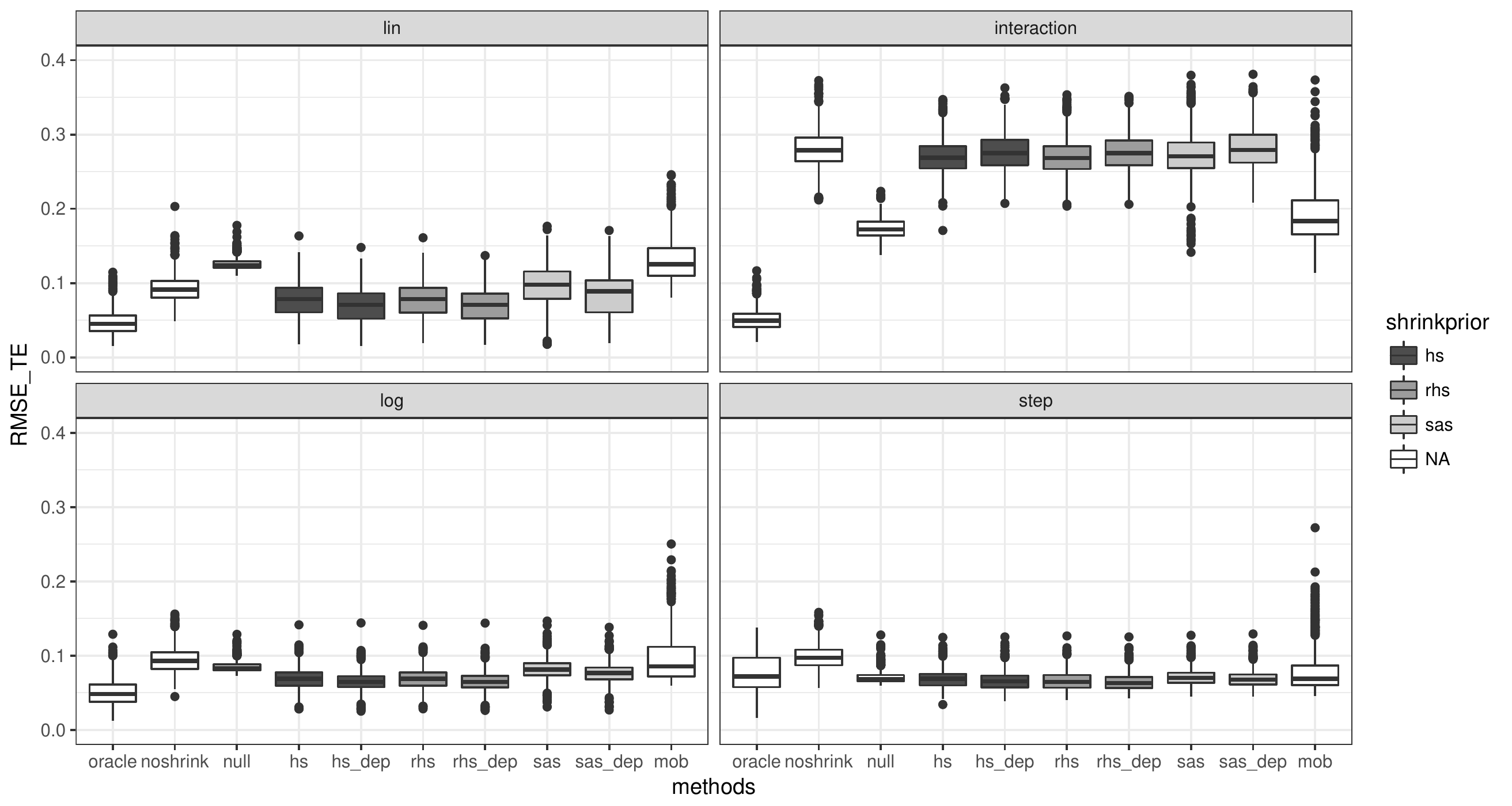}
\caption{Distribution of RMSE over 1000 simulated trials for treatment effect predictions averaged over patients and doses.
with $n=500$ (100 per group) and $k=10$ covariates for different
functional forms in scenarios with prognostic and predictive covariates.}
\label{fig:rmse_nonlin_n500_k10}
\end{figure}
\begin{table}[h]
\caption{Subgroup identification metrics for scenarios with $n=500$ (100 per group) and $k=10$ covariates for different
functional forms in scenarios with prognostic and predictive covariates.}
\centering
\scriptsize
\begin{tabular}{llrrrrrrr}\toprule
functional form & method & $|S|$ & $|\hat{S}|$ & non.null & sens & spec & ppv & npv \\ 
  \midrule
linear & oracle & 175.29 & 150.00 & 1.00 & 0.78 & 0.96 & 0.93 & 0.90 \\ 
  & noshrink & 175.29 & 107.94 & 1.00 & 0.51 & 0.94 & 0.84 & 0.79 \\ 
  & null & 175.29 & 42.50 & 0.09 & 0.09 & 0.92 & 0.36 & 0.65 \\ 
  & hs & 175.29 & 138.05 & 0.93 & 0.62 & 0.91 & 0.84 & 0.84 \\ 
  & hs\_dep & 175.29 & 149.93 & 0.98 & 0.68 & 0.91 & 0.84 & 0.86 \\ 
  & rhs & 175.29 & 139.04 & 0.93 & 0.62 & 0.91 & 0.84 & 0.84 \\ 
  & rhs\_dep & 175.29 & 149.99 & 0.98 & 0.68 & 0.91 & 0.84 & 0.86 \\ 
  & sas & 175.29 & 118.19 & 0.80 & 0.50 & 0.91 & 0.80 & 0.79 \\ 
  & sas\_dep & 175.29 & 132.30 & 0.86 & 0.58 & 0.91 & 0.81 & 0.82 \\ 
  & mob & 175.29 & 178.85 & 0.93 & 0.60 & 0.77 & 0.64 & 0.80 \\ 
\midrule 
interaction & oracle & 203.49 & 169.93 & 1.00 & 0.76 & 0.95 & 0.93 & 0.86 \\ 
  & noshrink & 203.49 & 115.49 & 1.00 & 0.33 & 0.83 & 0.58 & 0.64 \\ 
  & null & 203.49 & 147.50 & 0.29 & 0.29 & 0.70 & 0.41 & 0.59 \\ 
  & hs & 203.49 & 139.31 & 1.00 & 0.40 & 0.81 & 0.60 & 0.66 \\ 
  & hs\_dep & 203.49 & 140.42 & 1.00 & 0.40 & 0.80 & 0.59 & 0.66 \\ 
  & rhs & 203.49 & 141.43 & 1.00 & 0.41 & 0.81 & 0.60 & 0.67 \\ 
  & rhs\_dep & 203.49 & 142.05 & 1.00 & 0.42 & 0.81 & 0.60 & 0.67 \\ 
  & sas & 203.49 & 148.80 & 0.99 & 0.41 & 0.78 & 0.57 & 0.66 \\ 
  & sas\_dep & 203.49 & 146.77 & 1.00 & 0.40 & 0.78 & 0.57 & 0.66 \\ 
  & mob & 203.49 & 201.92 & 0.87 & 0.55 & 0.70 & 0.62 & 0.73 \\ 
\midrule
logistic & oracle & 139.85 & 119.49 & 0.92 & 0.67 & 0.93 & 0.85 & 0.89 \\ 
 & noshrink & 139.85 & 74.13 & 1.00 & 0.33 & 0.92 & 0.64 & 0.78 \\ 
  & null & 139.85 & 18.00 & 0.04 & 0.04 & 0.96 & 0.29 & 0.72 \\ 
  & hs & 139.85 & 104.61 & 0.77 & 0.44 & 0.88 & 0.67 & 0.82 \\ 
  & hs\_dep & 139.85 & 100.95 & 0.82 & 0.44 & 0.89 & 0.70 & 0.82 \\ 
  & rhs & 139.85 & 97.97 & 0.75 & 0.41 & 0.89 & 0.68 & 0.82 \\ 
  & rhs\_dep & 139.85 & 96.38 & 0.82 & 0.42 & 0.90 & 0.70 & 0.82 \\ 
  & sas & 139.85 & 101.88 & 0.68 & 0.39 & 0.87 & 0.57 & 0.81 \\ 
  & sas\_dep & 139.85 & 82.34 & 0.60 & 0.32 & 0.90 & 0.60 & 0.79 \\ 
  & mob & 139.85 & 164.16 & 0.74 & 0.54 & 0.75 & 0.48 & 0.84 \\ 
\midrule
step function & oracle & 124.90 & 100.54 & 0.60 & 0.58 & 0.92 & 0.79 & 0.89 \\ 
 & noshrink & 124.90 & 65.50 & 0.98 & 0.26 & 0.91 & 0.50 & 0.79 \\ 
 & null & 124.90 & 16.50 & 0.03 & 0.03 & 0.97 & 0.25 & 0.75 \\ 
 & hs & 124.90 & 109.97 & 0.76 & 0.45 & 0.86 & 0.56 & 0.85 \\ 
 & hs\_dep & 124.90 & 94.02 & 0.71 & 0.38 & 0.88 & 0.58 & 0.83 \\ 
 & rhs & 124.90 & 99.50 & 0.72 & 0.40 & 0.87 & 0.57 & 0.84 \\ 
 & rhs\_dep & 124.90 & 84.96 & 0.70 & 0.34 & 0.89 & 0.58 & 0.82 \\ 
 & sas & 124.90 & 110.14 & 0.74 & 0.41 & 0.84 & 0.48 & 0.83 \\ 
 & sas\_dep & 124.90 & 87.09 & 0.58 & 0.32 & 0.87 & 0.49 & 0.82 \\ 
 & mob & 124.90 & 127.23 & 0.53 & 0.44 & 0.81 & 0.44 & 0.85 \\\bottomrule 
\end{tabular}
\label{tab:subid_nonlin_n100_k10}
\end{table}
\FloatBarrier
\bibliographystyle{wileyj}

\bibliography{bibl}

\newpage

\newpage 
\begin{center}
\textbf{\huge Supporting Material}
\end{center}

\setcounter{section}{0}
\setcounter{figure}{0}
\setcounter{table}{0}

In the following additional simulations results from the simulation study in Section 4 of the main article
for scenarios with smaller sample size ($n = 250$) or larger number of covariates ($k = 30$) are depicted. For details
on the setup of the simulation study and the metrics used see the main article.

\section{Estimation of individual treatment effect curves}

\begin{figure}[h!]
\centering
\includegraphics[width=\textwidth]{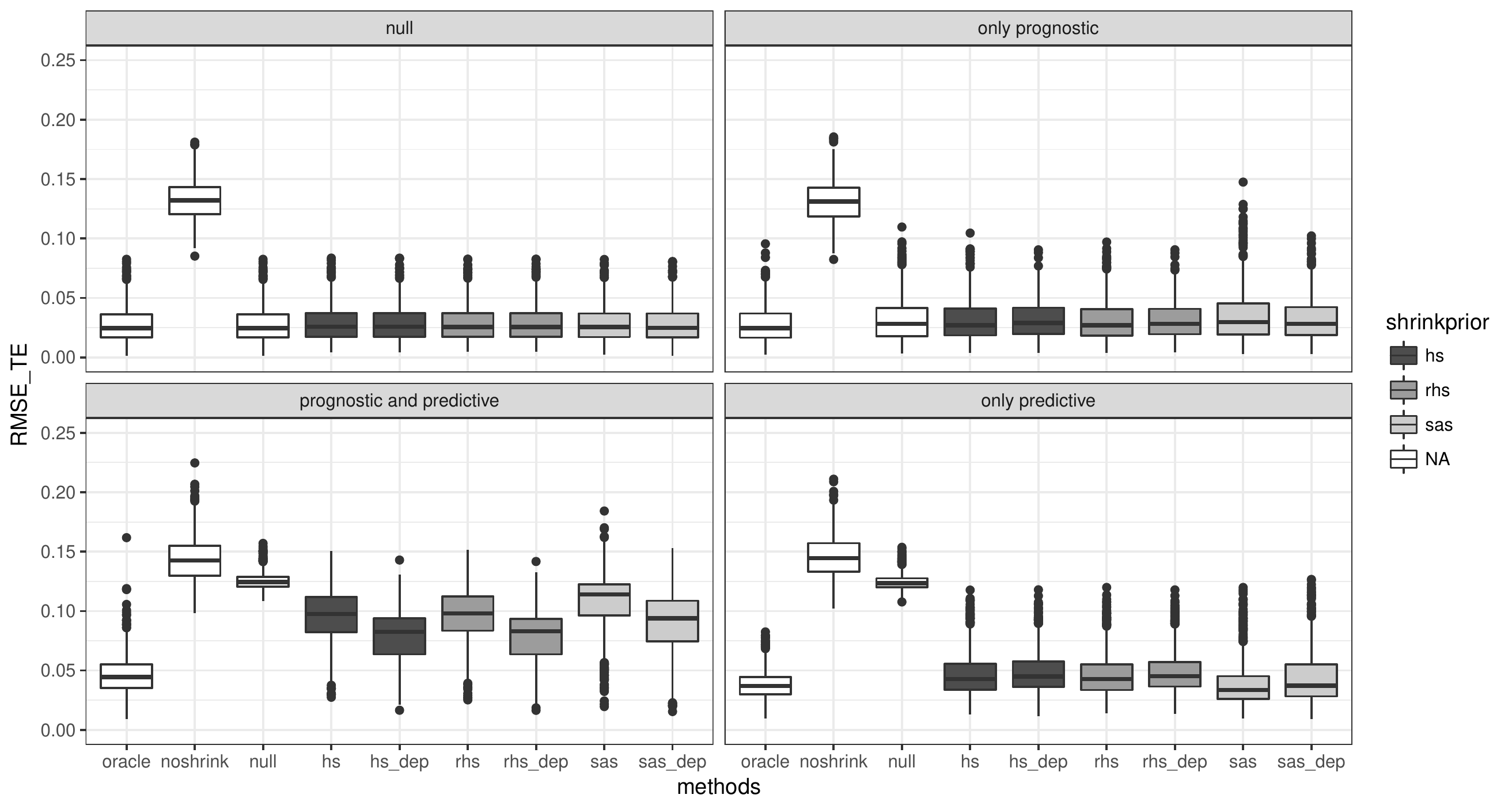}
\caption{Distribution of RMSE over 1000 simulated trials for treatment effect predictions averaged over patients and doses.
with $n=500$ (100 per group) and $k=30$ covariates}
\label{fig:rmse_n500_k30}
\end{figure}
\begin{figure}[h!]
\centering
\includegraphics[width=\textwidth]{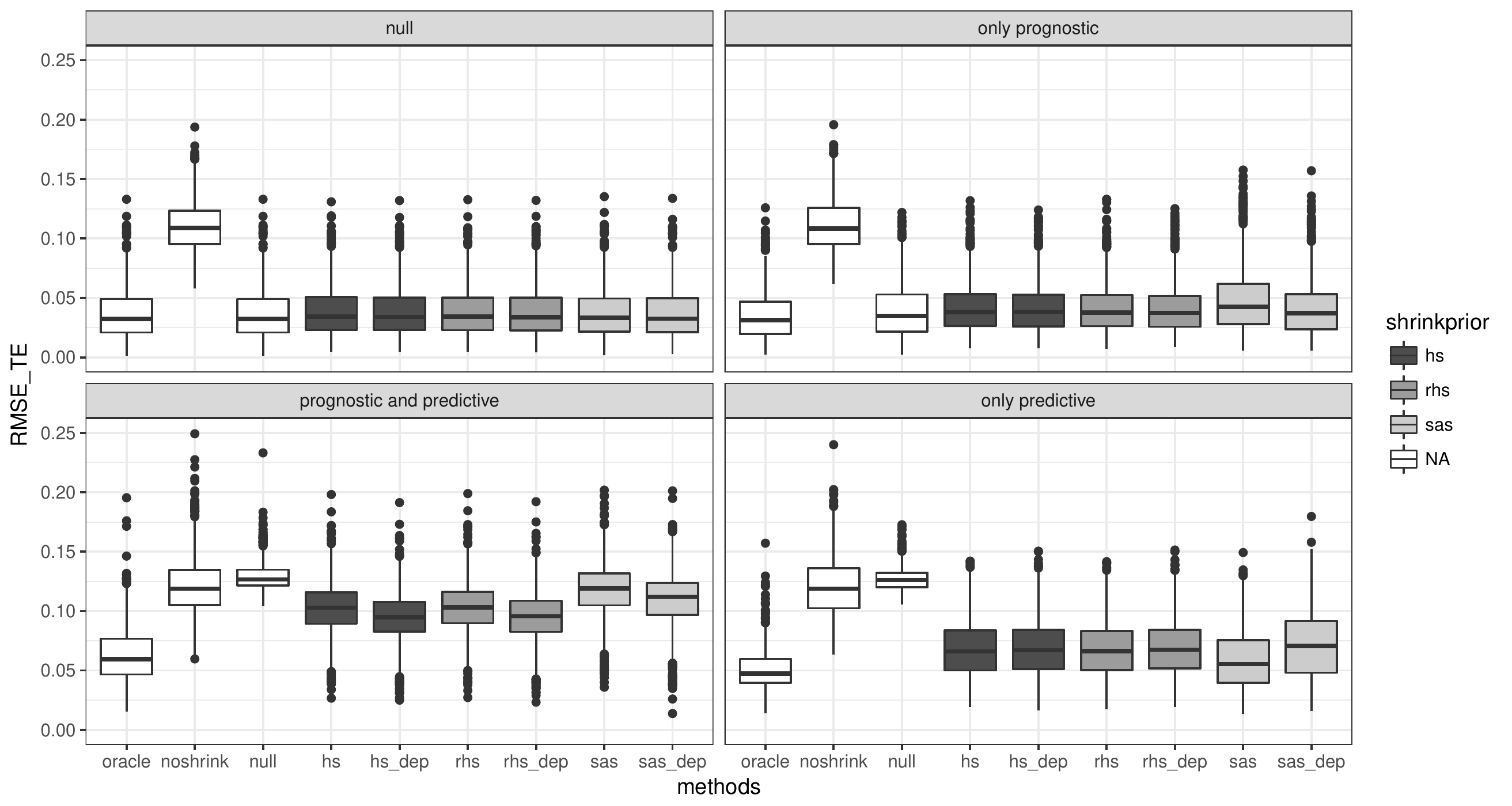}
\caption{Distribution of RMSE over 1000 simulated trials for treatment effect predictions averaged over patients and doses
with $n=250$ (50 per group) and $k=10$ covariates}
\label{fig:rmse_n250_k10}
\end{figure}
\begin{figure}[h!]
\centering
\includegraphics[width=\textwidth]{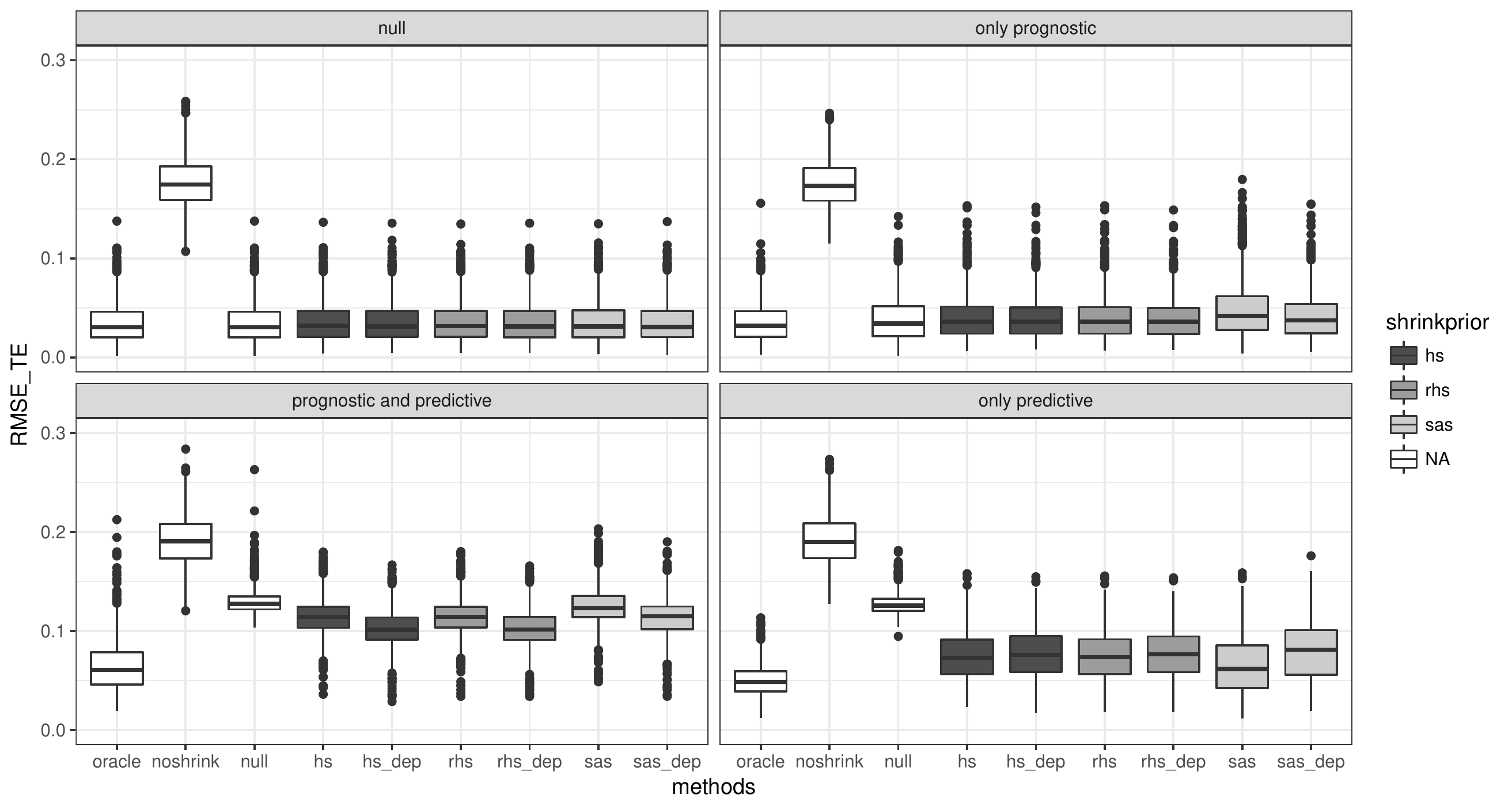}
\caption{Distribution of RMSE over 1000 simulated trials for treatment effect predictions averaged over patients and doses
with $n=250$ (50 per group) and $k=30$ covariates}
\end{figure}

\FloatBarrier

\section{Selection of predictive covariates}
\begin{table}[ht]
\caption{Rel. frequency of covariates being selected as having effects on $E_0$, $E_{max}$ and $ED_{50}$ with $n=250$ (50 per group) and $k=10$ covariates.
Variables selected with \textit{oracle} are the correct covariates.}
\scriptsize
\centering
\begin{tabular}{ll|rrrr|rrrr|rrrr}\hline
& & \multicolumn{4}{c}{$E_0$} & \multicolumn{4}{c}{$ED_{50}$} & \multicolumn{4}{c}{$E_{max}$}\\
scenario & method & x1 & x2 & x3 & x4 & x1 & x2 & x3 & x4 & x1 & x2 & x3 & x4 \\ 
  \hline
null & oracle & 0 & 0 & 0 & 0 & 0 & 0 & 0 & 0 & 0 & 0 & 0 & 0  \\  
  & noshrink & 0.52 & 0.53 & 0.54 & 0.53 & 0.47 & 0.47 & 0.47 & 0.49 & 0.46 & 0.45 & 0.47 & 0.48 \\ 
  & hs & 0.00 & 0.00 & 0.00 & 0.00 & 0.00 & 0.00 & 0.00 & 0.00 & 0.00 & 0.00 & 0.00 & 0.00 \\ 
  & hs\_dep & 0.00 & 0.00 & 0.00 & 0.00 & 0.00 & 0.00 & 0.00 & 0.00 & 0.00 & 0.00 & 0.00 & 0.00 \\ 
  & rhs & 0.00 & 0.00 & 0.00 & 0.00 & 0.00 & 0.00 & 0.00 & 0.00 & 0.00 & 0.00 & 0.00 & 0.00 \\ 
  & rhs\_dep & 0.00 & 0.00 & 0.00 & 0.00 & 0.00 & 0.00 & 0.00 & 0.00 & 0.00 & 0.00 & 0.00 & 0.00 \\ 
  & sas & 0.00 & 0.00 & 0.00 & 0.00 & 0.00 & 0.00 & 0.00 & 0.00 & 0.00 & 0.00 & 0.00 & 0.00 \\ 
  & sas\_dep & 0.00 & 0.00 & 0.00 & 0.00 & 0.00 & 0.00 & 0.00 & 0.00 & 0.00 & 0.00 & 0.00 & 0.00 \\ 
\hline 
only progn. & oracle & 1 & 1 & 1 & 0 & 0 & 0 & 0 & 0 & 0 & 0 & 0 & 0  \\
  & noshrink & 1.00 & 1.00 & 0.90 & 0.52 & 0.46 & 0.46 & 0.47 & 0.47 & 0.51 & 0.46 & 0.47 & 0.48 \\ 
  & hs & 0.98 & 0.98 & 0.65 & 0.02 & 0.00 & 0.00 & 0.00 & 0.00 & 0.01 & 0.01 & 0.01 & 0.00 \\ 
  & hs\_dep & 0.99 & 0.99 & 0.81 & 0.06 & 0.00 & 0.00 & 0.00 & 0.00 & 0.01 & 0.01 & 0.01 & 0.00 \\ 
  & rhs & 0.98 & 0.98 & 0.66 & 0.02 & 0.00 & 0.00 & 0.00 & 0.00 & 0.01 & 0.01 & 0.01 & 0.00 \\ 
  & rhs\_dep & 0.99 & 0.99 & 0.80 & 0.06 & 0.00 & 0.00 & 0.00 & 0.00 & 0.01 & 0.01 & 0.00 & 0.00 \\ 
  & sas & 0.93 & 0.92 & 0.29 & 0.00 & 0.01 & 0.01 & 0.01 & 0.00 & 0.05 & 0.05 & 0.03 & 0.00 \\ 
  & sas\_dep & 0.97 & 0.98 & 0.36 & 0.00 & 0.01 & 0.00 & 0.01 & 0.00 & 0.01 & 0.01 & 0.01 & 0.00 \\  
\hline
progn. \& pred. & oracle & 1 & 1 & 1 & 0 & 0 & 1 & 1 & 0 & 0 & 1 & 1 & 0  \\  
  & noshrink & 1.00 & 1.00 & 0.72 & 0.53 & 0.47 & 0.73 & 0.74 & 0.50 & 0.51 & 0.84 & 0.81 & 0.48 \\ 
  & hs & 0.95 & 0.80 & 0.02 & 0.01 & 0.00 & 0.03 & 0.01 & 0.00 & 0.03 & 0.24 & 0.05 & 0.00 \\ 
  & hs\_dep & 0.97 & 0.97 & 0.09 & 0.03 & 0.00 & 0.04 & 0.01 & 0.00 & 0.02 & 0.33 & 0.05 & 0.00 \\ 
  & rhs & 0.95 & 0.78 & 0.02 & 0.01 & 0.00 & 0.03 & 0.01 & 0.00 & 0.03 & 0.25 & 0.04 & 0.00 \\ 
  & rhs\_dep & 0.97 & 0.97 & 0.08 & 0.03 & 0.00 & 0.04 & 0.01 & 0.00 & 0.03 & 0.36 & 0.05 & 0.00 \\ 
  & sas & 0.89 & 0.68 & 0.01 & 0.00 & 0.02 & 0.12 & 0.03 & 0.00 & 0.08 & 0.32 & 0.06 & 0.00 \\ 
  & sas\_dep & 0.96 & 0.80 & 0.03 & 0.00 & 0.01 & 0.12 & 0.02 & 0.00 & 0.03 & 0.27 & 0.05 & 0.00 \\ 
\hline
only pred. & oracle & 0 & 0 & 0 & 0 & 0 & 1 & 1 & 0 & 0 & 1 & 1 & 0  \\ 
  & noshrink & 0.57 & 0.66 & 0.66 & 0.54 & 0.47 & 0.73 & 0.73 & 0.46 & 0.45 & 0.83 & 0.83 & 0.48 \\ 
  & hs & 0.00 & 0.11 & 0.11 & 0.00 & 0.00 & 0.07 & 0.09 & 0.00 & 0.00 & 0.51 & 0.51 & 0.00 \\ 
  & hs\_dep & 0.00 & 0.14 & 0.13 & 0.00 & 0.00 & 0.07 & 0.08 & 0.00 & 0.00 & 0.47 & 0.48 & 0.00 \\ 
  & rhs & 0.00 & 0.12 & 0.11 & 0.00 & 0.00 & 0.09 & 0.10 & 0.00 & 0.01 & 0.53 & 0.54 & 0.01 \\ 
  & rhs\_dep & 0.00 & 0.14 & 0.14 & 0.00 & 0.00 & 0.07 & 0.08 & 0.00 & 0.00 & 0.49 & 0.50 & 0.00 \\ 
  & sas & 0.00 & 0.07 & 0.06 & 0.00 & 0.00 & 0.42 & 0.41 & 0.00 & 0.00 & 0.72 & 0.73 & 0.00 \\ 
  & sas\_dep & 0.00 & 0.16 & 0.18 & 0.00 & 0.00 & 0.27 & 0.27 & 0.00 & 0.00 & 0.52 & 0.53 & 0.00 \\\hline 
\end{tabular}
\label{tab:varsec_50_10}
\end{table}

\begin{table}[ht]
\caption{Rel. frequency of covariates being selected as having effects on $E_0$, $E_{max}$ and $ED_{50}$ with $n=500$ (100 per group) and $k=30$ covariates.
Variables selected with \textit{oracle} are the correct covariates.}
\scriptsize
\centering
\begin{tabular}{ll|rrrr|rrrr|rrrr}\hline
& & \multicolumn{4}{c}{$E_0$} & \multicolumn{4}{c}{$ED_{50}$} & \multicolumn{4}{c}{$E_{max}$}\\
scenario & method & x1 & x2 & x3 & x4 & x1 & x2 & x3 & x4 & x1 & x2 & x3 & x4 \\ 
  \hline
null & oracle & 0 & 0 & 0 & 0 & 0 & 0 & 0 & 0 & 0 & 0 & 0 & 0  \\  
  & noshrink & 0.53 & 0.55 & 0.54 & 0.55 & 0.48 & 0.47 & 0.49 & 0.46 & 0.53 & 0.49 & 0.50 & 0.53 \\ 
  & hs & 0.00 & 0.00 & 0.00 & 0.00 & 0.00 & 0.00 & 0.00 & 0.00 & 0.00 & 0.00 & 0.00 & 0.00 \\ 
  & hs\_dep & 0.00 & 0.00 & 0.00 & 0.00 & 0.00 & 0.00 & 0.00 & 0.00 & 0.00 & 0.00 & 0.00 & 0.00 \\ 
  & rhs & 0.00 & 0.00 & 0.00 & 0.00 & 0.00 & 0.00 & 0.00 & 0.00 & 0.00 & 0.00 & 0.00 & 0.00 \\ 
  & rhs\_dep & 0.00 & 0.00 & 0.00 & 0.00 & 0.00 & 0.00 & 0.00 & 0.00 & 0.00 & 0.00 & 0.00 & 0.00 \\ 
  & sas & 0.00 & 0.00 & 0.00 & 0.00 & 0.00 & 0.00 & 0.00 & 0.00 & 0.00 & 0.00 & 0.00 & 0.00 \\ 
  & sas\_dep & 0.00 & 0.00 & 0.00 & 0.00 & 0.00 & 0.00 & 0.00 & 0.00 & 0.00 & 0.00 & 0.00 & 0.00 \\ 
\hline 
only progn. & oracle & 1 & 1 & 1 & 0 & 0 & 0 & 0 & 0 & 0 & 0 & 0 & 0  \\
  & noshrink & 1.00 & 1.00 & 0.97 & 0.53 & 0.45 & 0.45 & 0.47 & 0.48 & 0.52 & 0.49 & 0.51 & 0.52 \\ 
  & hs & 0.99 & 0.99 & 0.84 & 0.00 & 0.00 & 0.00 & 0.01 & 0.00 & 0.01 & 0.01 & 0.01 & 0.00 \\ 
  & hs\_dep & 1.00 & 1.00 & 0.90 & 0.01 & 0.01 & 0.00 & 0.00 & 0.00 & 0.00 & 0.00 & 0.00 & 0.00 \\ 
  & rhs & 0.99 & 0.99 & 0.84 & 0.00 & 0.00 & 0.00 & 0.01 & 0.00 & 0.01 & 0.01 & 0.01 & 0.00 \\ 
  & rhs\_dep & 1.00 & 1.00 & 0.90 & 0.00 & 0.00 & 0.00 & 0.00 & 0.00 & 0.00 & 0.00 & 0.00 & 0.00 \\ 
  & sas & 0.98 & 0.98 & 0.63 & 0.00 & 0.01 & 0.01 & 0.03 & 0.00 & 0.01 & 0.02 & 0.05 & 0.00 \\ 
  & sas\_dep & 0.99 & 0.99 & 0.69 & 0.00 & 0.00 & 0.00 & 0.01 & 0.00 & 0.01 & 0.01 & 0.01 & 0.00 \\
\hline
progn. \& pred. & oracle & 1 & 1 & 1 & 0 & 0 & 1 & 1 & 0 & 0 & 1 & 1 & 0  \\   
  & noshrink & 1.00 & 1.00 & 0.75 & 0.53 & 0.48 & 0.78 & 0.80 & 0.46 & 0.50 & 0.92 & 0.89 & 0.50 \\ 
  & hs & 0.99 & 0.88 & 0.02 & 0.00 & 0.00 & 0.09 & 0.02 & 0.00 & 0.01 & 0.36 & 0.13 & 0.00 \\ 
  & hs\_dep & 1.00 & 1.00 & 0.16 & 0.00 & 0.01 & 0.18 & 0.05 & 0.00 & 0.01 & 0.69 & 0.19 & 0.00 \\ 
  & rhs & 0.99 & 0.87 & 0.02 & 0.00 & 0.00 & 0.10 & 0.03 & 0.00 & 0.01 & 0.37 & 0.12 & 0.00 \\ 
  & rhs\_dep & 1.00 & 1.00 & 0.15 & 0.00 & 0.01 & 0.20 & 0.07 & 0.00 & 0.02 & 0.70 & 0.18 & 0.00 \\ 
  & sas & 0.97 & 0.76 & 0.02 & 0.00 & 0.01 & 0.21 & 0.07 & 0.00 & 0.03 & 0.36 & 0.12 & 0.00 \\ 
  & sas\_dep & 0.99 & 0.96 & 0.13 & 0.00 & 0.01 & 0.34 & 0.12 & 0.00 & 0.01 & 0.56 & 0.18 & 0.00 \\ 
\hline
only pred. & oracle & 0 & 0 & 0 & 0 & 0 & 1 & 1 & 0 & 0 & 1 & 1 & 0  \\   
  & noshrink & 0.54 & 0.79 & 0.78 & 0.54 & 0.49 & 0.77 & 0.78 & 0.49 & 0.51 & 0.90 & 0.88 & 0.51 \\ 
  & hs & 0.00 & 0.04 & 0.06 & 0.00 & 0.00 & 0.26 & 0.28 & 0.00 & 0.00 & 0.87 & 0.84 & 0.00 \\ 
  & hs\_dep & 0.00 & 0.07 & 0.07 & 0.00 & 0.00 & 0.24 & 0.24 & 0.00 & 0.00 & 0.84 & 0.83 & 0.00 \\ 
  & rhs & 0.00 & 0.04 & 0.06 & 0.00 & 0.00 & 0.27 & 0.28 & 0.00 & 0.00 & 0.88 & 0.86 & 0.00 \\ 
  & rhs\_dep & 0.00 & 0.06 & 0.07 & 0.00 & 0.00 & 0.24 & 0.25 & 0.00 & 0.00 & 0.86 & 0.84 & 0.00 \\ 
  & sas & 0.00 & 0.04 & 0.04 & 0.00 & 0.00 & 0.74 & 0.77 & 0.00 & 0.00 & 0.94 & 0.94 & 0.00 \\ 
  & sas\_dep & 0.00 & 0.08 & 0.10 & 0.00 & 0.00 & 0.64 & 0.64 & 0.00 & 0.00 & 0.88 & 0.85 & 0.00 \\\hline 
\end{tabular}
\label{tab:varsec_100_30}
\end{table}

\begin{table}[ht]
\caption{Rel. frequency of covariates being selected as having effects on $E_0$, $E_{max}$ and $ED_{50}$ with $n=250$ (50 per group) and $k=30$ covariates.
Variables selected with \textit{oracle} are the correct covariates.}
\scriptsize
\centering
\begin{tabular}{ll|rrrr|rrrr|rrrr}\hline
& & \multicolumn{4}{c}{$E_0$} & \multicolumn{4}{c}{$ED_{50}$} & \multicolumn{4}{c}{$E_{max}$}\\
scenario & method & x1 & x2 & x3 & x4 & x1 & x2 & x3 & x4 & x1 & x2 & x3 & x4 \\ 
  \hline
null & oracle & 0 & 0 & 0 & 0 & 0 & 0 & 0 & 0 & 0 & 0 & 0 & 0  \\ 
  & noshrink & 0.54 & 0.54 & 0.52 & 0.57 & 0.39 & 0.39 & 0.42 & 0.39 & 0.50 & 0.48 & 0.51 & 0.50 \\ 
  & hs & 0.00 & 0.00 & 0.00 & 0.00 & 0.00 & 0.00 & 0.00 & 0.00 & 0.00 & 0.00 & 0.00 & 0.00 \\ 
  & hs\_dep & 0.00 & 0.00 & 0.00 & 0.00 & 0.00 & 0.00 & 0.00 & 0.00 & 0.00 & 0.00 & 0.00 & 0.00 \\ 
  & rhs & 0.00 & 0.00 & 0.00 & 0.00 & 0.00 & 0.00 & 0.00 & 0.00 & 0.00 & 0.00 & 0.00 & 0.00 \\ 
  & rhs\_dep & 0.00 & 0.00 & 0.00 & 0.00 & 0.00 & 0.00 & 0.00 & 0.00 & 0.00 & 0.00 & 0.00 & 0.00 \\ 
  & sas & 0.00 & 0.00 & 0.00 & 0.00 & 0.00 & 0.00 & 0.00 & 0.00 & 0.00 & 0.00 & 0.00 & 0.00 \\ 
  & sas\_dep & 0.00 & 0.00 & 0.00 & 0.00 & 0.00 & 0.00 & 0.00 & 0.00 & 0.00 & 0.00 & 0.00 & 0.00 \\ 
\hline 
only progn. & oracle & 1 & 1 & 1 & 0 & 0 & 0 & 0 & 0 & 0 & 0 & 0 & 0  \\
  & noshrink & 1.00 & 1.00 & 0.89 & 0.56 & 0.40 & 0.39 & 0.42 & 0.39 & 0.47 & 0.51 & 0.49 & 0.51 \\ 
  & hs & 0.98 & 0.98 & 0.46 & 0.01 & 0.00 & 0.00 & 0.00 & 0.00 & 0.01 & 0.01 & 0.00 & 0.00 \\ 
  & hs\_dep & 0.99 & 1.00 & 0.52 & 0.00 & 0.00 & 0.00 & 0.00 & 0.00 & 0.00 & 0.00 & 0.00 & 0.00 \\ 
  & rhs & 0.98 & 0.98 & 0.46 & 0.01 & 0.00 & 0.00 & 0.00 & 0.00 & 0.01 & 0.01 & 0.00 & 0.00 \\ 
  & rhs\_dep & 0.99 & 1.00 & 0.51 & 0.00 & 0.00 & 0.00 & 0.00 & 0.00 & 0.00 & 0.00 & 0.00 & 0.00 \\ 
  & sas & 0.92 & 0.91 & 0.24 & 0.00 & 0.02 & 0.03 & 0.01 & 0.00 & 0.05 & 0.06 & 0.02 & 0.00 \\ 
  & sas\_dep & 0.97 & 0.97 & 0.29 & 0.00 & 0.01 & 0.00 & 0.00 & 0.00 & 0.01 & 0.01 & 0.00 & 0.00 \\ 
\hline
progn. \& pred. & oracle & 1 & 1 & 1 & 0 & 0 & 1 & 1 & 0 & 0 & 1 & 1 & 0  \\ 
  & noshrink & 1.00 & 1.00 & 0.62 & 0.55 & 0.43 & 0.55 & 0.55 & 0.38 & 0.51 & 0.75 & 0.75 & 0.50 \\ 
  & hs & 0.95 & 0.76 & 0.00 & 0.00 & 0.00 & 0.02 & 0.00 & 0.00 & 0.02 & 0.21 & 0.01 & 0.00 \\ 
  & hs\_dep & 0.97 & 0.98 & 0.02 & 0.00 & 0.00 & 0.04 & 0.01 & 0.00 & 0.01 & 0.28 & 0.02 & 0.00 \\ 
  & rhs & 0.96 & 0.75 & 0.00 & 0.00 & 0.00 & 0.03 & 0.01 & 0.00 & 0.02 & 0.22 & 0.01 & 0.00 \\ 
  & rhs\_dep & 0.97 & 0.98 & 0.02 & 0.00 & 0.00 & 0.05 & 0.01 & 0.00 & 0.02 & 0.31 & 0.02 & 0.00 \\ 
  & sas & 0.91 & 0.68 & 0.00 & 0.00 & 0.02 & 0.15 & 0.02 & 0.00 & 0.06 & 0.31 & 0.03 & 0.00 \\ 
  & sas\_dep & 0.96 & 0.81 & 0.02 & 0.00 & 0.00 & 0.12 & 0.01 & 0.00 & 0.02 & 0.27 & 0.02 & 0.00 \\ 
\hline
only pred. & oracle & 0 & 0 & 0 & 0 & 0 & 1 & 1 & 0 & 0 & 1 & 1 & 0  \\  
  & noshrink & 0.53 & 0.76 & 0.75 & 0.57 & 0.41 & 0.55 & 0.56 & 0.42 & 0.51 & 0.73 & 0.74 & 0.51 \\ 
  & hs & 0.00 & 0.11 & 0.12 & 0.00 & 0.00 & 0.06 & 0.07 & 0.00 & 0.00 & 0.43 & 0.41 & 0.00 \\ 
  & hs\_dep & 0.00 & 0.15 & 0.15 & 0.00 & 0.00 & 0.05 & 0.06 & 0.00 & 0.00 & 0.36 & 0.34 & 0.00 \\ 
  & rhs & 0.00 & 0.11 & 0.12 & 0.00 & 0.00 & 0.07 & 0.08 & 0.00 & 0.00 & 0.46 & 0.42 & 0.00 \\ 
  & rhs\_dep & 0.00 & 0.16 & 0.15 & 0.00 & 0.00 & 0.05 & 0.06 & 0.00 & 0.00 & 0.37 & 0.35 & 0.00 \\ 
  & sas & 0.00 & 0.07 & 0.06 & 0.00 & 0.00 & 0.36 & 0.39 & 0.00 & 0.00 & 0.65 & 0.65 & 0.00 \\ 
  & sas\_dep & 0.00 & 0.14 & 0.16 & 0.00 & 0.00 & 0.21 & 0.22 & 0.00 & 0.00 & 0.44 & 0.41 & 0.00 \\\hline 
\end{tabular}
\label{tab:varsec_50_30}
\end{table}

\FloatBarrier

\section{Subgroup identification}
\begin{table}[ht]
\caption{Subgroup identification metrics for scenarios with $n=250$ (50 per group) and $k=10$ covariates.}
\scriptsize
\centering
\begin{tabular}{ll|rrrrrrr}\hline
scenario & method & $|S|$ & $|\hat{S}|$ & non.null & sens & spec & ppv & npv \\ 
  \hline
null & oracle & 0.00 & 14.75 & 0.06 &  & 0.94 & 0.00 & 1.00 \\ 
  & noshrink & 0.00 & 30.48 & 0.98 &  & 0.88 & 0.00 & 1.00 \\ 
  & hs & 0.00 & 14.62 & 0.18 &  & 0.94 & 0.00 & 1.00 \\ 
  & hs\_dep & 0.00 & 15.45 & 0.18 &  & 0.94 & 0.00 & 1.00 \\ 
  & rhs & 0.00 & 15.08 & 0.19 &  & 0.94 & 0.00 & 1.00 \\ 
  & rhs\_dep & 0.00 & 15.32 & 0.18 &  & 0.94 & 0.00 & 1.00 \\ 
  & sas & 0.00 & 14.26 & 0.09 &  & 0.94 & 0.00 & 1.00 \\ 
  & sas\_dep & 0.00 & 14.78 & 0.07 &  & 0.94 & 0.00 & 1.00 \\
\hline 
only progn. & oracle & 0.00 & 14.75 & 0.06 &  & 0.94 & 0.00 & 1.00 \\ 
  & noshrink & 0.00 & 29.41 & 0.98 &  & 0.88 & 0.00 & 1.00 \\ 
  & hs & 0.00 & 21.89 & 0.34 &  & 0.91 & 0.00 & 1.00 \\ 
  & hs\_dep & 0.00 & 21.82 & 0.35 &  & 0.91 & 0.00 & 1.00 \\ 
  & rhs & 0.00 & 22.28 & 0.35 &  & 0.91 & 0.00 & 1.00 \\ 
  & rhs\_dep & 0.00 & 21.88 & 0.34 &  & 0.91 & 0.00 & 1.00 \\ 
  & sas & 0.00 & 27.76 & 0.41 &  & 0.89 & 0.00 & 1.00 \\ 
  & sas\_dep & 0.00 & 21.49 & 0.28 &  & 0.91 & 0.00 & 1.00 \\
\hline
progn. \& pred. & oracle & 87.87 & 68.82 & 0.99 & 0.68 & 0.95 & 0.91 & 0.86 \\ 
  & noshrink & 87.87 & 50.76 & 1.00 & 0.42 & 0.91 & 0.74 & 0.75 \\ 
  & hs & 87.87 & 63.77 & 0.82 & 0.48 & 0.87 & 0.76 & 0.78 \\ 
  & hs\_dep & 87.87 & 67.91 & 0.89 & 0.53 & 0.87 & 0.78 & 0.80 \\ 
  & rhs & 87.87 & 63.88 & 0.82 & 0.48 & 0.87 & 0.76 & 0.78 \\ 
  & rhs\_dep & 87.87 & 67.66 & 0.89 & 0.53 & 0.87 & 0.78 & 0.80 \\ 
  & sas & 87.87 & 62.92 & 0.71 & 0.42 & 0.84 & 0.68 & 0.75 \\ 
  & sas\_dep & 87.87 & 60.41 & 0.73 & 0.42 & 0.86 & 0.71 & 0.76 \\
\hline
only pred. & oracle & 87.27 & 71.26 & 1.00 & 0.75 & 0.97 & 0.95 & 0.89 \\ 
  & noshrink & 87.27 & 49.73 & 1.00 & 0.42 & 0.92 & 0.74 & 0.75 \\ 
  & hs & 87.27 & 74.21 & 0.97 & 0.68 & 0.91 & 0.87 & 0.87 \\ 
  & hs\_dep & 87.27 & 73.51 & 0.96 & 0.68 & 0.91 & 0.87 & 0.86 \\ 
  & rhs & 87.27 & 74.69 & 0.97 & 0.69 & 0.91 & 0.87 & 0.87 \\ 
  & rhs\_dep & 87.27 & 73.56 & 0.96 & 0.67 & 0.91 & 0.87 & 0.86 \\ 
  & sas & 87.27 & 77.80 & 0.99 & 0.75 & 0.92 & 0.88 & 0.89 \\ 
  & sas\_dep & 87.27 & 71.36 & 0.94 & 0.66 & 0.91 & 0.86 & 0.85 \\\hline 
\end{tabular}
\label{tab:subid_50_10}
\end{table}

\begin{table}[ht]
\caption{Subgroup identification metrics for scenarios with $n=500$ (100 per group) and $k=30$ covariates.}
\scriptsize
\centering
\begin{tabular}{ll|rrrrrrr}\hline
scenario & method & $|S|$ & $|\hat{S}|$ & non.null & sens & spec & ppv & npv \\ 
  \hline
null & oracle & 0.00 & 9.50 & 0.02 &  & 0.98 & 0.00 & 1.00 \\ 
  & noshrink & 0.00 & 70.34 & 1.00 &  & 0.86 & 0.00 & 1.00 \\ 
  & hs & 0.00 & 10.23 & 0.08 &  & 0.98 & 0.00 & 1.00 \\ 
  & hs\_dep & 0.00 & 10.82 & 0.07 &  & 0.98 & 0.00 & 1.00 \\ 
  & rhs & 0.00 & 10.20 & 0.08 &  & 0.98 & 0.00 & 1.00 \\ 
  & rhs\_dep & 0.00 & 10.50 & 0.07 &  & 0.98 & 0.00 & 1.00 \\ 
  & sas & 0.00 & 10.13 & 0.03 &  & 0.98 & 0.00 & 1.00 \\ 
  & sas\_dep & 0.00 & 9.41 & 0.02 &  & 0.98 & 0.00 & 1.00 \\ 
\hline 
only progn. & oracle & 0.00 & 13.33 & 0.03 &  & 0.97 & 0.00 & 1.00 \\ 
  & noshrink & 0.00 & 70.60 & 1.00 &  & 0.86 & 0.00 & 1.00 \\ 
  & hs & 0.00 & 20.09 & 0.16 &  & 0.96 & 0.00 & 1.00 \\ 
  & hs\_dep & 0.00 & 20.83 & 0.23 &  & 0.96 & 0.00 & 1.00 \\ 
  & rhs & 0.00 & 19.62 & 0.16 &  & 0.96 & 0.00 & 1.00 \\ 
  & rhs\_dep & 0.00 & 20.27 & 0.22 &  & 0.96 & 0.00 & 1.00 \\ 
  & sas & 0.00 & 28.67 & 0.22 &  & 0.94 & 0.00 & 1.00 \\ 
  & sas\_dep & 0.00 & 20.44 & 0.16 &  & 0.96 & 0.00 & 1.00 \\
\hline
progn. \& pred. & oracle & 175.91 & 147.92 & 1.00 & 0.77 & 0.96 & 0.94 & 0.90 \\ 
  & noshrink & 175.91 & 99.73 & 1.00 & 0.37 & 0.90 & 0.66 & 0.73 \\ 
  & hs & 175.91 & 110.30 & 0.80 & 0.45 & 0.91 & 0.81 & 0.78 \\ 
  & hs\_dep & 175.91 & 139.46 & 0.95 & 0.61 & 0.90 & 0.82 & 0.83 \\ 
  & rhs & 175.91 & 110.84 & 0.81 & 0.46 & 0.91 & 0.81 & 0.78 \\ 
  & rhs\_dep & 175.91 & 139.53 & 0.96 & 0.61 & 0.90 & 0.82 & 0.83 \\ 
  & sas & 175.91 & 104.31 & 0.66 & 0.39 & 0.89 & 0.74 & 0.75 \\ 
  & sas\_dep & 175.91 & 123.99 & 0.84 & 0.52 & 0.90 & 0.79 & 0.80 \\ 
\hline
only pred. & oracle & 87.27 & 71.26 & 1.00 & 0.75 & 0.97 & 0.95 & 0.89 \\ 
  & noshrink & 87.27 & 49.73 & 1.00 & 0.42 & 0.92 & 0.74 & 0.75 \\ 
  & hs & 87.27 & 74.21 & 0.97 & 0.68 & 0.91 & 0.87 & 0.87 \\ 
  & hs\_dep & 87.27 & 73.51 & 0.96 & 0.68 & 0.91 & 0.87 & 0.86 \\ 
  & rhs & 87.27 & 74.69 & 0.97 & 0.69 & 0.91 & 0.87 & 0.87 \\ 
  & rhs\_dep & 87.27 & 73.56 & 0.96 & 0.67 & 0.91 & 0.87 & 0.86 \\ 
  & sas & 87.27 & 77.80 & 0.99 & 0.75 & 0.92 & 0.88 & 0.89 \\ 
  & sas\_dep & 87.27 & 71.36 & 0.94 & 0.66 & 0.91 & 0.86 & 0.85 \\\hline 
\end{tabular}
\label{tab:subid_100_30}
\end{table}

\begin{table}[ht]
\caption{Subgroup identification metrics for scenarios with $n=250$ (50 per group) and $k=30$ covariates.}
\scriptsize
\centering
\begin{tabular}{ll|rrrrrrr}\hline
scenario & method & $|S|$ & $|\hat{S}|$ & non.null & sens & spec & ppv & npv \\ 
  \hline
null & oracle & 0.00 & 13.25 & 0.05 &  & 0.95 & 0.00 & 1.00 \\ 
  & noshrink & 0.00 & 42.36 & 1.00 &  & 0.83 & 0.00 & 1.00 \\ 
  & hs & 0.00 & 14.30 & 0.12 &  & 0.94 & 0.00 & 1.00 \\ 
  & hs\_dep & 0.00 & 14.37 & 0.12 &  & 0.94 & 0.00 & 1.00 \\ 
  & rhs & 0.00 & 14.21 & 0.13 &  & 0.94 & 0.00 & 1.00 \\ 
  & rhs\_dep & 0.00 & 14.32 & 0.12 &  & 0.94 & 0.00 & 1.00 \\ 
  & sas & 0.00 & 15.12 & 0.08 &  & 0.94 & 0.00 & 1.00 \\ 
  & sas\_dep & 0.00 & 13.85 & 0.07 &  & 0.94 & 0.00 & 1.00 \\ 
\hline 
only progn. & oracle & 0.00 & 15.75 & 0.06 &  & 0.94 & 0.00 & 1.00 \\ 
  & noshrink & 0.00 & 43.22 & 1.00 &  & 0.83 & 0.00 & 1.00 \\ 
  & hs & 0.00 & 21.37 & 0.28 &  & 0.91 & 0.00 & 1.00 \\ 
  & hs\_dep & 0.00 & 21.46 & 0.28 &  & 0.91 & 0.00 & 1.00 \\ 
  & rhs & 0.00 & 20.80 & 0.27 &  & 0.92 & 0.00 & 1.00 \\ 
  & rhs\_dep & 0.00 & 21.17 & 0.28 &  & 0.92 & 0.00 & 1.00 \\ 
  & sas & 0.00 & 29.95 & 0.38 &  & 0.88 & 0.00 & 1.00 \\ 
  & sas\_dep & 0.00 & 22.39 & 0.26 &  & 0.91 & 0.00 & 1.00 \\
\hline
progn. and pred. & oracle & 87.81 & 67.24 & 0.99 & 0.67 & 0.95 & 0.91 & 0.86 \\ 
  & noshrink & 87.81 & 51.95 & 1.00 & 0.32 & 0.86 & 0.55 & 0.70 \\ 
  & hs & 87.81 & 54.20 & 0.65 & 0.37 & 0.87 & 0.71 & 0.75 \\ 
  & hs\_con & 87.81 & 61.66 & 0.80 & 0.45 & 0.87 & 0.75 & 0.77 \\ 
  & rhs & 87.81 & 54.16 & 0.66 & 0.37 & 0.87 & 0.71 & 0.75 \\ 
  & rhs\_con & 87.81 & 61.37 & 0.81 & 0.45 & 0.87 & 0.76 & 0.77 \\ 
  & sas & 87.81 & 57.26 & 0.62 & 0.37 & 0.85 & 0.65 & 0.74 \\ 
  & sas\_con & 87.81 & 55.47 & 0.67 & 0.38 & 0.86 & 0.70 & 0.75 \\ 
\hline
 only pred. & oracle & 87.73 & 71.61 & 1.00 & 0.75 & 0.96 & 0.95 & 0.89 \\ 
  & noshrink & 87.73 & 52.25 & 1.00 & 0.33 & 0.85 & 0.55 & 0.70 \\ 
  & hs & 87.73 & 74.12 & 0.93 & 0.65 & 0.90 & 0.84 & 0.85 \\ 
  & hs\_dep & 87.73 & 72.64 & 0.92 & 0.64 & 0.90 & 0.85 & 0.85 \\ 
  & rhs & 87.73 & 74.50 & 0.94 & 0.65 & 0.90 & 0.84 & 0.85 \\ 
  & rhs\_dep & 87.73 & 72.37 & 0.92 & 0.63 & 0.90 & 0.84 & 0.85 \\ 
  & sas & 87.73 & 78.09 & 0.96 & 0.72 & 0.91 & 0.86 & 0.88 \\ 
  & sas\_dep & 87.73 & 69.65 & 0.89 & 0.61 & 0.90 & 0.84 & 0.84 \\\hline 
\end{tabular}
\label{tab:subid_50_30}
\end{table}

\end{document}